\documentclass[useAMS, usenatbib, usegraphicx]{mn2e} 

\title[Intermediate-mass black holes] 
{Intermediate-Mass Black Hole Growth and Feedback in Dwarf Galaxies at High Redshifts} 

\author[P. Barai et al.] 
{Paramita Barai$^{1}$, 
Elisabete M. de Gouveia Dal Pino$^{1}$ 
\vspace{0.2cm} \\ 
$^{1}$ Instituto de Astronomia, Geof\'isica e Ci\^encias Atmosf\'ericas - 
Universidade de S\~ao Paulo (IAG-USP), \\ 
Rua do Mat\~ao 1226, S\~ao Paulo, 05508-090, Brasil 
}

\begin{document} 

\maketitle 


\begin{abstract} 

Intermediate-mass black holes (IMBHs: masses between $100 - 10^{6} M_{\odot}$) historically 
comprise of an elusive population compared to stellar-mass and supermassive BHs. 
Recently IMBHs have started to be observed at the centers of low-mass galaxies. 
We perform cosmological hydrodynamical simulations of $(2 h^{-1} ~ {\rm Mpc})^3$ 
comoving boxes and investigate the growth and feedback of 
central IMBHs in dwarf galaxies (DGs). 
The earliest BHs appear at $z \sim 18 - 25$, and grow thereafter 
by accreting gas and by merger with other BHs. 
We find that, starting from $10^{2} M_{\odot}$, it is possible to build up IMBHs of 
a few$\times 10^{5} - 10^{6} M_{\odot}$ by $z = 5$, when the BHs 
are seeded in halos less massive than $4 \times 10^{7} M_{\odot}$. 
The BH accretion rates increase with time, and reaches 
$\dot{M}_{\rm BH} = (0.2 - 0.8) \dot{M}_{\rm Edd}$ for the massive IMBHs by $z = 4$. 
The star formation rate density (SFRD) evolution of the DGs 
(stellar mass $10^{5} - 10^{8} M_{\odot}$) has a peak plateau between $z = 4 - 6$. 
Star formation is quenched between $z = 9 - 4$. 
The SFRD is reduced by factors up to $3$, 
when the BHs have grown to a few times $10^5 M_{\odot}$. 
Even in the presence of stronger SN-driven mass ejection, 
the BHs continue to grow up to $z \sim 6$, sustained by gas inflows 
driven by galaxy mergers and interactions in a cosmological environment. 
Our conclusions, based on numerical simulation results, 
support the scenario that early feedback from IMBHs in gas-rich DGs at $z = 5 - 8$ 
can potentially solve several anomalies in the DG mass range within the 
concordance $\Lambda$CDM cosmological scenario \citep{Silk17}. 
Our results suggest that IMBHs at DG centers grow faster than their host galaxies 
in the early Universe, and the resulting BH feedback turns the DGs and the BHs dormant. 

\end{abstract} 



\section{Introduction} 
\label{sec-intro} 

Black holes (BHs) are usually observed to be of stellar-mass 
$(M_{\rm BH} \sim 1 - 100 M_{\odot})$, 
or supermassive $(M_{\rm BH} \sim 10^{6} - 10^{9} M_{\odot})$. 
Stellar-mass BHs have been historically observed in X-ray binaries existing in our 
Milky Way and other galaxies \citep[e.g.,][]{Bolton72, Cowley83, Orosz07, Corral16}, 
and more recently in globular clusters \citep[e.g.,][]{Maccarone07, Giesers18}. 
Supermassive BHs (SMBHs) are believed to exist at the centers of 
active galactic nuclei (AGN), which liberate enormous amounts of feedback energy 
powered by the accretion of matter \citep[e.g.,][]{Rees84, Ferrarese05}. 
AGN are widely observed through their multi-wavelength emission 
at all cosmic epochs, $z = 0 - 7$, starting from the local Universe 
up to $13$ Gyr ago \citep[e.g.,][]{Urry95, Goulding09, Mortlock11}. 


Intermediate-mass black holes (IMBHs) of masses between $100 - 10^{6} M_{\odot}$ 
\citep[e.g.,][]{vanderMarel04, Graham18} are however not as widely observed. 
Some studies argue that the accreting BHs in ultra-luminous X-ray sources 
could be IMBHs \citep[e.g.,][]{Miller03, Sutton12, Caballero-Garcia18, Mezcua18b}, 
based on the calculation of the BH mass from a cold thermal disc 
accreting at sub-Eddington rates. 


We do not know how the SMBHs in AGN grew to billions of solar masses. 
In one scenario, the SMBHs could grow from stellar-mass BHs 
by rapid enhanced accretion (e.g. super-Eddington growth). 
Or, some exotic processes in the early Universe could have formed massive BH seeds 
(in the direct collapse scenario) soon after the Big Bang. 
Massive seed IMBHs could be a possible explanation for the origin of SMBHs. 
A recent study by \citet{Kovetz18} suggest that gravitational wave measurements using 
the advanced LIGO over six years can be used to limit the formation mechanism of 
IMBHs by the runaway merger of stellar-mass BHs in globular clusters. 

Relatively recently, massive BHs have started to be observed 
hosted in Dwarf Galaxies (DGs). 
Dynamical BH mass limits detected in nearby dwarfs include several examples such as: 
$M_{\rm BH} \sim 3 \times 10^{6} M_{\odot}$ in the 
dwarf elliptical galaxy M32 \citep{vanderMarel98}; 
$M_{\rm BH} \sim 10^{5} M_{\odot}$ in the 
dwarf lenticular field galaxy NGC 404 \citep{Seth10}; 
$M_{\rm BH} < 2 \times 10^{4} M_{\odot}$ in the 
dwarf elliptical galaxy NGC 205 \citep{Valluri05}; and 
$M_{\rm BH} < 10^{6} M_{\odot}$ in the dwarf spheroidal galaxy Ursa Minor \citep{Demers95}. 

Some of the central IMBHs in DGs show signatures of activity 
in the form of low-luminosity AGN. 
NGC 4395 is a bulgeless dwarf galaxy with an extremely faint 
Seyfert 1 nucleus \citep{Filippenko89}, and an estimated central BH mass 
$M_{\rm BH} \sim 10^{4} - 3 \times 10^{5} M_{\odot}$ \citep{Peterson05, Woo19}. 
Pox 52 (G 1200-2038) is a dwarf galaxy with Seyfert characteristics \citep{Kunth87}, 
having a central BH of 
$M_{\rm BH} \sim (2 - 4) \times 10^{5} M_{\odot}$ \citep{Thornton08}. 
The dwarf starburst galaxy Henize 2-10 \citep{Reines11} has an actively accreting 
massive BH with an order-of-magnitude mass $M_{\rm BH} \sim 10^{6} M_{\odot}$. 

The population of DGs with observed AGN signatures have been increasing 
\citep[e.g.,][]{Chilingarian18}. 
\citet{Izotov08} presented the spectra of $4$ low-metallicity DGs with very-high 
broad H$\alpha$ luminosities, most likely coming from accretion disks around IMBHs 
($5 \times 10^{5} \leq M_{\rm BH} \leq 3 \times 10^{6} M_{\odot}$). 
Performing a systematic search for AGN in DGs using optical spectroscopy 
from the SDSS, \citet{Reines13} found $136$ DGs, with stellar mass between 
$10^{8.5} < M_{\star} < 10^{9.5} M_{\odot}$ at $z < 0.055$, hosting active massive 
BHs with virial BH masses in the range $10^{5} < M_{\rm BH} < 10^{6} M_{\odot}$. 
Using SDSS DR7, \citet{Moran14} identified $28$ nearby low-mass, low-luminosity 
DGs containing accreting IMBHs ($10^{3} \leq M_{\rm BH} \leq 10^{6} M_{\odot}$) 
at their centers, and derived a lower limit of a few percent 
on the fraction of DGs containing AGN. 


The BH mass versus stellar mass relationship of IMBHs in DGs tend to be 
below the existing local relation extending into the lower galaxy mass 
regime \citep[e.g.,][]{Baldassare15, Reines15}. 
Albeit there is a large scatter, and exceptions i.e. 
IMBHs lying on the $[M_{\rm BH} - \sigma_{\star}]$ local relation 
when extended to lower masses \citep[e.g.,][]{Woo19}. 
IMBH candidates detected using deep X-ray observations 
\citep[e.g.,][]{Schramm13, Secrest15, Lemons15} show that 
low-mass galaxies with $M_{\star} < 3 \times 10^{9} M_{\odot}$ have BHs of 
$M_{\rm BH} \sim 2 \times 10^{5} M_{\odot}$. 
The most-massive ultracompact dwarf galaxy M59-UCD3 
with $M_{\star} \sim 2 \times 10^{8} M_{\odot}$ has a massive central BH 
of $M_{\rm BH} \sim 4 \times 10^{6} M_{\odot}$ \citep{Ahn18}. 
The Fornax UCD3 hosting a central $M_{\rm BH} \sim 3.3 \times 10^{6} M_{\odot}$ 
corresponds to $4 \%$ of the galaxy stellar mass \citep{Afanasiev18}. 
\citet{Marleau17} investigated the presence of AGN in nearby galaxies 
using mid-infrared emission, and found some DGs in their sample. 
These DGs have stellar masses between $10^{6} - 10^{9} M_{\odot}$, 
and the BH masses (estimated from their IR luminosity, assuming a 
bolometric luminosity of $10 \%$ of the Eddington) 
in the range $10^{3} - 10^{6} M_{\odot}$. 

Recently the highest-redshift discovery of AGN in DGs has been made by \citet{Mezcua18a}. 
They present a sample of 40 AGN of type 2 at $z \leq 2.4$ with luminosities 
in the range $L_{0.5 - 10 {\rm keV}} \sim 10^{39} - 10^{44}$ erg/s. 
The hosts are DGs with stellar masses between $10^{7} - 3 \times 10^{9} M_{\odot}$, 
selected from the Chandra COSMOS-Legacy survey. 
Estimating the BH masses using the local scaling relation between BH 
and stellar mass \citep{Reines15}, all the AGN are consistent with 
hosting $\sim 10^{4} - 10^{5} M_{\odot}$ IMBHs (with a scatter of $0.55$ dex), 
of typical Eddington ratios 
(or, the ratio of the BH accretion rate to the Eddington rate) $> 1\%$. 
According to \citet{Mezcua18a}, the observational trends suggest that AGN in DGs 
evolve differently than those in more-massive galaxies. 

AGN influence the formation and evolution of galaxies 
in the form of feedback, affecting the environment from pc to Mpc scales 
\citep[e.g.,][]{Silk98, Barai08, Fabian12}. 
Supermassive BHs and host galaxies are argued to coevolve, generating 
observational trends such as the central BH mass - host galaxy stellar 
bulge mass correlations \citep[e.g.,][]{Magorrian98, Gebhardt00}. 
The SMBH energy output is often observed as various forms of AGN outflows 
\citep[e.g.,][]{Crenshaw03, Cicone14, Melioli15, Tombesi15, Barai18}. 

Analogous to SMBHs producing AGN feedback, the IMBHs are also expected to have 
feedback: the energy radiated by IMBHs should affect their host galaxies, 
possibly driving galactic outflows. 
BH or AGN feedback mechanism has recently started to be observed in low-mass galaxies. 
\citet{Penny17} presented observational evidence for AGN feedback in a sample of 
$69$ quenched low-mass galaxies $(M_{\star} < 4 \times 10^{9} M_{\odot})$; 
including $6$ galaxies showing signatures of an active AGN 
preventing ongoing star-formation. 
At the same time, supernova feedback should also be operating parallelly 
in these low-mass galaxies, ejecting out gas \citep[e.g.,][]{Caproni17}. 

AGN feedback operates mostly in the negative form quenching star formation, 
as indicated by simulations \citep[e.g.,][]{Scannapieco05, Barai14}, 
and some observations \citep[e.g.,][]{Schawinski06, Lanz16}. 
At the same time, AGN outflows can occasionally compress clumpy gas clouds 
and trigger starbursts, bringing in positive AGN feedback, 
as found in numerical studies \citep[e.g.,][]{DeYoung89, Bieri15, Mukherjee18}, 
and observed in jet-induced star formation 
\citep[e.g.,][]{Chambers87, Viegas92, Zinn13, Salome17}. 
In this work we focus on negative BH feedback effects where star-formation is quenched. 

The concordance $\Lambda$CDM cosmological scenario of galaxy formation presents 
multiple anomalies in the dwarf galaxy mass range: e.g. core-cusp, number of DGs. 
Recently \citet{Silk17} made an exciting theoretical claim that the presence of IMBHs 
at the centers of essentially all old DGs can potentially solve the problems. 
Early feedback energy from these IMBHs can affect the host 
gas-rich dwarf galaxies at $z = 5 - 8$. 
This early feedback can quench star-formation, reduce the number of DGs, 
and impact the density profile at DG centers. 
\citet{Dashyan17} studied the same problem analytically using a 
constant luminosity of the AGN, and compared AGN versus supernova (SN) feedback. 
They find a critical halo mass 
below which the central AGN can drive gas out of the host halo. 
This negative feedback effect of AGN is found to be more efficient than SN 
in the most-massive DGs, where SN is not able to expel the gas. 
Noteworthy that some numerical simulations find that 
AGN feedback is inefficient in low-mass galaxies 
\citep{Dubois15, Habouzit17, Angles-Alcazar17, Prieto17, Trebitsch18}, 
because strong SN feedback hampers BH growth by removing gas from around the BH. 
Such a scenario is also suppoted by observational studies 
\citep{Martin-Navarro18} indicating that AGN activity does not 
dominate baryonic cooling in low-mass galaxies where SN feedback prevails. 

In this work, we investigate the scenario that IMBHs are present at the centers 
of dwarf galaxies, by performing cosmological hydrodynamical simulations. 
Our goals are to (i) test if IMBHs would grow at DG centers in a cosmological 
environment, and (ii) quantify the impact of feedback from IMBHs on their host DGs, 
especially the effects on star formation at cosmic epochs $z = 6 - 4$. 

This paper is organised as follows: 
our numerical code and simulation setup are described in \S\ref{sec-numerical}, 
our results are presented and analyzed in \S\ref{sec-results}, 
and a summary of our main findings are given in \S\ref{sec-conclusion}.

\section{Numerical Method} 
\label{sec-numerical} 

We use a modified version of the code {\sc GADGET-3} \citep{Springel05}, 
which uses the TreePM (particle mesh) and SPH (smoothed particle hydrodynamics) methods. 
It includes sub-resolution physics as described in \S\ref{sec-num-cool-SF-SN} 
and \S\ref{sec-num-BH}. 
The physics of multiphase structure of the interstellar medium (ISM), on scales 
unresolved in cosmological simulations, is modeled using spatially averaged 
properties describing the medium on scales that are resolved. 
Our different simulation runs are outlined in \S\ref{sec-num-Sim}.

\subsection{Cooling, Star-Formation, SN Feedback}    
\label{sec-num-cool-SF-SN} 

Radiative cooling and heating is implemented, 
including metal-line cooling \citep{Wiersma09a}. 
In this model, net cooling rates are computed element-by-element 
tracking $11$ atomic species: H, He, C, Ca, O, N, Ne, Mg, S, Si, Fe. 
A spatially-uniform time-dependent photoionizing background radiation is 
considered from the cosmic microwave background and the 
\citet{Haardt01} model for the ultraviolet/X-ray background. 
The gas is assumed to be dust free, optically thin, 
and in (photo-) ionization equilibrium. 
Contributions from the $11$ elements are interpolated as a function of density, 
temperature and redshift from tables that have been pre-computed using the 
public photoionization code CLOUDY \citep[last described by][]{Ferland98}. 

Star formation (SF) is adopted following the multiphase 
effective sub-resolution model of \citet{SH03}. 
Gas particles with density above a limiting threshold, 
$\rho_{\rm SF} = 0.13$ cm$^{-3}$ (in units of number density of hydrogen atoms), 
represent cold and hot phase regions of the ISM. 
The unresolved cold condensed clouds (comprising of fraction $x$ 
of a gas particle) are in pressure equilibrium with the ambient hot gas. 
The star formation rate of the high-density gas particle (of mass $m_{\rm gas}$) 
is estimated using a Schmidt-type law \citep{Schmidt59}: 
$\dot{m}_{\star} = x m_{\rm gas} / t_{\star}$. 
Here, $x m_{\rm gas}$ is the mass of cold clouds supplying the material 
available for SF. The SF time-scale ($t_{\star}$) of the gas particle 
is computed as a function of gas density: 
$t_{\star} (\rho) = t_0^{\star} (\rho / \rho_{\rm SF})^{-1/2}$, 
where $t_0^{\star}$ is a parameter of the model. 
We use a value of $t_0^{\star} = 1.5$ Gyr \citep{SH03, Tornatore07}, 
chosen to reproduce the observed relation between the disc-averaged 
SF per unit area and the gas surface density \citep{Kennicutt98}. 
The star forming gas remains in self-regulated equilibrium. 
Star particles are collisionless, and are spawned from these high-density 
gas particles, according to a stochastic scheme \citep{Katz96}. 

Stellar evolution and chemical enrichment are computed for the 11 elements 
\citep{Tornatore07}. 
Each star particle is treated as a simple stellar population (SSP). 
Given a stellar initial mass function (IMF), the mass of the SSP is varied in time 
following the death of stars, and accounting for stellar mass losses. 
A fixed stellar initial mass function \citep{Chabrier03} is included, 
in the mass range $(0.1 - 100) M_{\odot}$. 
Stars within a mass interval $[8 - 40] M_{\odot}$ become SN first before 
turning into stellar-mass black holes at the end of their lives, 
while stars of mass $> 40 M_{\odot}$ are allowed to 
directly end in black holes without contributing to gas enrichment. 

Feedback from supernovae is incorporated in the kinetic form, 
assuming a mass ejection rate $\dot{M}_{\rm SN}$ 
proportional to the star formation rate ($\dot{M}_{\star}$): 
\begin{equation} 
\dot{M}_{\rm SN} = \eta \dot{M}_{\star} . 
\end{equation} 
The mass loading factor of SN wind is taken as $\eta = 2$ 
\citep[e.g.,][]{Tornatore07, Barai13, Melioli13}, 
following observations revealing that SN-driven outflow rates are comparable to 
or larger than SF rates of galaxies \citep[e.g.,][]{Martin99, Bouche12}. 
The SN wind kinetic power is a fixed fraction $\chi$ of SN internal energy rate: 
\begin{equation} 
\label{eq-SN-feedback} 
\frac{1}{2} \dot{M}_{\rm SN} v_{\rm SN}^2 = \chi \epsilon_{SN} \dot{M}_{\star} . 
\end{equation} 
Here $v_{\rm SN}$ is the SN wind velocity, 
$\epsilon_{SN}$ is the average energy released by SN for each $M_{\odot}$ 
of stars formed under the instantaneous recycling approximation. 
For our adopted \citet{Chabrier03} power-law IMF, 
$\epsilon_{SN} = 1.1 \times 10^{49}$ erg $M_{\odot}^{-1}$. 
Combining above expressions, $v_{\rm SN}$ can be re-written as: 
$v_{\rm SN} = \left( 2 \chi \epsilon_{SN} / \eta \right)^{1/2}$. 
Following a series of studies \citep[e.g.,][]{Tornatore07, Tescari11, Barai13}, 
and unlike \citep{SH03}, we choose $v_{\rm SN}$ as a free parameter. 
We adopt a constant-velocity outflow with SN wind velocity 
$v_{\rm SN} = 350$ km/s \citep[as was done in e.g.][]{Tornatore07, Barai15, Biffi16}. 

The numerical implementation of launching SN wind in the {\sc GADGET-3} code 
involves a probabilistic criteria similar to spawning star particles 
\citep[following e.g.,][]{SH03, Barai13}. 
It assumes there is no time delay between star formation and SN explosions. 
A probability is calculated in a timestep $\Delta t$ for each 
high-density multiphase gas particle: 
\begin{equation} 
p = 1 - \exp \left( - \frac{\eta x \Delta t}{t_{\star}} \right). 
\end{equation} 
A random number is drawn in the interval [0, 1], 
and if it is below $p$ then the gas particle is given a SN wind kick. 
If $\vec{v}$ is particle velocity and $\phi$ its gravitational potential, 
then its velocity is updated to: 
$\vec{v}_{\rm new} = \vec{v}_{\rm old} + v_{\rm SN} \hat{y}$. 
The direction (unit vector $\hat{y}$) is set along positive or negative 
$(\vec{v}_{\rm old} \times \vec{\nabla} \phi)$, 
which makes the SN wind particles to be preferentially ejected along the 
rotation axis of the galaxy or perpendicular to the galaxy disk. 

In order to enable the SN outflows to escape from dense SF regions without 
affecting the SF, the new SN wind particle is decoupled from hydrodynamic 
interactions, for a maximum duration $t_{\rm dec} = 0.025 t_{\rm H}(z)$, 
where $t_{\rm H}(z)$ is the Hubble time at a relevant simulation redshift. 
Then the SN wind particle does not enter in the hydrodynamic force computation 
in the code, but it is included in the gravity and SPH density calculations. 
If the gas particle's density has fallen below 
$\rho_{\rm dec} = 0.25 \rho_{\rm SF}$, then the decoupling is stopped before 
$t_{\rm dec}$, and full hydrodynamics is enabled again.

%

\begin{table*} 
\begin{minipage}{1.0 \linewidth} 
\caption{ 
Simulation runs and parameters. 
} 
\label{Table-Sims} 
\centering 
\begin{tabular}{@{}cccccc} 

\hline 

Run  & SN Wind Feedback         & BH      & Min. Halo Mass for BH Seeding, & Seed BH Mass,                & BH kinetic feedback \\ 
name & Mass Load Factor, $\eta$ & present & $M_{\rm HaloMin} [M_{\odot}]$  & $M_{\rm BHseed} [M_{\odot}]$ & kick velocity $v_w$ (km/s) \\ 

\hline 

{\it noSN-noBH}       & --  & No  & -- & -- & --                           \\ 

{\it SN}              & $2$ & No  & -- & -- & --                           \\ 

{\it BHs2h1e6v2}      & $2$ & Yes & $1 \times 10^{6}$ & $10^{2}$ &  $2000$ \\ 

{\it BHs2h1e6v10}     & $2$ & Yes & $1 \times 10^{6}$ & $10^{2}$ & $10000$ \\ 

{\it BHs2h4e7v2}      & $2$ & Yes & $4 \times 10^{7}$ & $10^{2}$ &  $2000$ \\ 

{\it BHs2h7e7v2}      & $2$ & Yes & $7 \times 10^{7}$ & $10^{2}$ &  $2000$ \\ 

{\it BHs3h1e7v2}      & $2$ & Yes & $1 \times 10^{7}$ & $10^{3}$ &  $2000$ \\ 

{\it BHs3h1e7v2-SNhi} & $5$ & Yes & $1 \times 10^{7}$ & $10^{3}$ &  $2000$ \\ 

{\it BHs3h2e7v2}      & $2$ & Yes & $2 \times 10^{7}$ & $10^{3}$ &  $2000$ \\ 

{\it BHs3h3e7v2}      & $2$ & Yes & $3 \times 10^{7}$ & $10^{3}$ &  $2000$ \\ 

{\it BHs3h3e7v1}      & $2$ & Yes & $3 \times 10^{7}$ & $10^{3}$ &  $1000$ \\ 

{\it BHs3h4e7v1}      & $2$ & Yes & $4 \times 10^{7}$ & $10^{3}$ &  $1000$ \\ 

{\it BHs3h4e7v2}      & $2$ & Yes & $4 \times 10^{7}$ & $10^{3}$ &  $2000$ \\ 

{\it BHs3h4e7v5}      & $2$ & Yes & $4 \times 10^{7}$ & $10^{3}$ &  $5000$ \\ 

{\it BHs3h4e7v10}     & $2$ & Yes & $4 \times 10^{7}$ & $10^{3}$ & $10000$ \\ 

{\it BHs3h5e7v2}      & $2$ & Yes & $5 \times 10^{7}$ & $10^{3}$ &  $2000$ \\ 

{\it BHs4h4e7v2}      & $2$ & Yes & $4 \times 10^{7}$ & $10^{4}$ &  $2000$ \\ 


\hline 
\end{tabular} 

\end{minipage} 
\end{table*} 


\subsection{BH Accretion and Feedback} 
\label{sec-num-BH} 

BHs are collisionless sink particles (of mass $M_{\rm BH}$) in our simulations. 
A BH (of initial mass $M_{\rm BHseed}$) is seeded at the center of each halo more 
massive than a total mass $M_{\rm HaloMin}$, which does not contain a BH yet. 
Halos are identified by executing a {\it Friends-of-Friends} (FOF) 
group finder on-the-fly within our simulations. 
We test different values of minimum halo mass and seed BH mass in the range: 
$M_{\rm HaloMin} = (10^{6} - 5 \times 10^{7}) M_{\odot}$, 
and $M_{\rm BHseed} = (10^{2} - 10^{4}) M_{\odot}$. 

Gas is considered to accrete onto a BH according to the Bondi-Hoyle-Lyttleton 
accretion rate \citep[$\dot{M}_{\rm Bondi}$:][]{Hoyle39, Bondi52}, 
\begin{equation} 
\label{eq-Mdot-Bondi} 
\dot{M}_{\rm Bondi} = \alpha \frac{4 \pi G^2 M_{\rm BH}^2 \rho}{ \left(c_{s}^2 + v^2\right) ^ {3/2}} , 
\end{equation} 
where $G$ is the gravitational constant, $c_{s}$ is the sound speed, 
$\rho$ is the gas density, $v$ is the velocity of the BH relative to the gas, and 
$\alpha = 100$ is a numerical boost factor \citep[e.g.,][]{SDH05, Johansson09a, Dubois13}. 
Furthermore, accretion is limited to the Eddington mass accretion rate $(\dot{M}_{\rm Edd})$: 
$\dot{M}_{\rm BH} = {\rm min} \left( \dot{M}_{\rm Bondi}, \dot{M}_{\rm Edd} \right)$. 
The Eddington luminosity is used to express the Eddington mass accretion rate, 
\begin{equation} 
\label{eq-LEdd} 
L_{\rm Edd} = \frac{4 \pi G M_{\rm BH} m_p c} {\sigma_T} = \epsilon_r \dot{M}_{\rm Edd} c^2 , 
\end{equation} 
where $m_p$ is the mass of a proton, $c$ is the speed of light, 
and $\sigma_T$ is the Thomson scattering cross-section for an electron. 

The implementation in the {\sc GADGET-3} code involves computing 
a BH smoothing length $h_{\rm BH}$. 
It is determined by the implicit solution of the following equation at each timestep, 
\begin{equation} 
\label{eq-BH-Smooth} 
\frac{4}{3} \pi h_{\rm BH}^3 \rho_{\rm BH} = M_{\rm ngb} , 
\end{equation} 
where $\rho_{\rm BH}$ is the density at the position of the BH, 
and $M_{\rm ngb}$ is the mass of $200$ neighboring gas particles. 
A maximum of $100 L_{\rm soft}$ (with $L_{\rm soft} =$ gravitational 
softening length, viz. \S2.1 of \citet{Springel05}) 
is set for the possible value of $h_{\rm BH}$. 
A sphere of radius $h_{\rm BH}$ is used to compute accretion onto the BH 
of neighboring gas particles, using a stochastic methodology at every timestep 
\citep[for details see \S2.1 and \S2.3 of ][]{Barai14}. 
Physical quantities (e.g. $\rho$, $c_{s}$, $v$ in Eq.~(\ref{eq-Mdot-Bondi})) 
are computed dynamically within the code, 
using the SPH kernel-weighted smoothing method over gas particles 
neighboring every BH particle. 

Feedback energy is distributed to the surrounding gas, according to: 
\begin{equation} 
\label{eq-Edot-Feed} 
\dot{E}_{\rm feed} = \epsilon_f \epsilon_r \dot{M}_{\rm BH} c^2. 
\end{equation} 
Here $\epsilon_r$ is the radiative efficiency, 
and $\epsilon_f$ is the feedback efficiency. 
We adopt $\epsilon_r = 0.1$, which assumes radiatively efficient accretion 
onto a Schwarzschild BH \citep{Shakura73}. 

Kinetic BH feedback is included \citep[][]{Barai14, Barai16}, where the 
neighboring gas is pushed outward with a velocity $v_w$ and mass outflow rate $\dot{M}_w$. 
Using the conservation of energy, 
$\frac{1}{2} \dot{M}_w v_w^2 = \dot{E}_{\rm feed}$, 
and Eq.~(\ref{eq-Edot-Feed}), 
the BH kinetic outflow rate can be written as, 
\begin{equation} 
\label{eq-MdotW-EDW} 
\dot{M}_w = 2 \epsilon_f \epsilon_r \dot{M}_{\rm BH} \frac{c^2}{v_w^2} . 
\end{equation} 
We use the values: $\epsilon_f = 0.05$, and the range $v_w = 1000, 2000, 5000, 10000$ 
km/s.\footnote{We notice that, although most of the adopted values of the AGN wind velocity 
in our models are smaller than those inferred from observations of AGN ultra-fast outflows 
\citep[e.g.,][and references therein]{Melioli15, Kraemer18}, the few tests we performed 
with more compatible values around $10000$ km/s did not reveal significant differences 
in the results (for more details see \S\ref{sec-results}).} 

%


The BH kinetic feedback energy is distributed to the gas within a distance
$h_{\rm BH}$ from the BH. 
A bi-cone volume is defined around the BH, 
of slant height $h_{\rm BH}$ and half-opening angle $45^{\circ}$. 
The cone-axis direction is randomly assigned during a BH seeding, 
and remains fixed henceforth for each BH. 
The total gas mass within the bi-cone $M_{\rm gas}^{\rm vicinity}$ is computed. 
A probability is calculated for the $i$'th gas particle inside the bi-cone, 
in a timestep $\Delta t$: 
\begin{equation} 
\label{eq-probKick} 
p_i = \frac{\dot{M}_w \Delta t} {M_{\rm gas}^{\rm vicinity}} , 
\end{equation} 
where $\dot{M}_w$ is the mass outflow rate obtained from Eq.~(\ref{eq-MdotW-EDW}). 
A random number $x_i$ is drawn uniformly distributed in the interval $[0, 1]$. 
If $x_i < p_i$, then the gas particle is imparted a velocity boost by AGN wind, such that: 
\begin{equation} 
\label{eq-vNew} 
\vec{v}_{\rm new} = \vec{v}_{\rm old} + v_w \hat{n} . 
\end{equation} 
The AGN wind direction $\hat{n}$ is considered radially outward from the BH. 




We incorporate a scheme for BH {\it pinning}\footnote{The BH seeds start as 
$10^{2} M_{\odot}$ and $10^{3} M_{\odot}$ in most of our simulations 
(Table~\ref{Table-Sims}), which is of the order of $1/100$th of the gas 
and DM particle's mass (\S\ref{sec-num-Sim}). 
Such low-mass BHs are susceptible to dynamical movements 
and wander away from galaxy centres \citep[as seen in e.g.,][]{Barai18}, 
by interactions with neighboring more-massive gas and DM particles. 
Therefore a reposition algorithm is often adopted in SPH simulations 
to keep the BH at the galaxy center.}, 
or {\it BH advection algorithm} \citep[also done in e.g.,][]{SDH05, Wurster13, Schaye15}. 
Each BH is repositioned at each time-step to the location of the 
minimum gravitational potential of its host galaxy \citep{Barai18}. 
However such an algorithm has the limitation that it might overestimate 
the accretion rate onto the BH, because of forcing it to be positioned 
near the highest gas density location at the galaxy center. 
Note that we do not consider any dynamical friction effects experienced by a BH 
as it orbits within its host galaxy, which is expected to become 
important at sub-kpc and smaller scales according to higher resolution simulations 
\citep[e.g.,][]{Tremmel15}. 
Such scales are generally lower than or of the order of the BH smoothing length 
($h_{\rm BH}$) in our simulations, and thus not resolved. 
Therefore the BH dynamics is effectively replaced by the numerical advection algorithm. 



We consider that when galaxies merge during hierarchical structure assembly, 
their hosted central BHs merge as well.
In the numerical algorithm, two BH particles are allowed to merge 
to form a single BH, when the distance between them is smaller than the 
smoothing length of either one and their relative velocity is 
below the local sound speed \citep[e.g.,][]{Sijacki07, DiMatteo12}. 
The final merged BH has a mass equal to the sum of the BH masses, 
and a velocity along the center of mass of the initial two merging BHs. 
To impart kinetic feedback energy, the merged BH retains the 
bi-cone axis direction of the more massive initial BH.

\subsection{Simulations} 
\label{sec-num-Sim} 

We perform cosmological hydrodynamical simulations of small-sized 
boxes to probe dwarf galaxies at high redshifts. 
The initial condition at $z = 100$ is generated using the 
{\sc MUSIC}\footnote{MUSIC - Multi-scale Initial Conditions for Cosmological 
Simulations: https://bitbucket.org/ohahn/music} software \citep{Hahn11}. 
The concordance flat $\Lambda$CDM cosmological model is used: 
$\Omega_{M, 0} = 0.3089, \Omega_{\Lambda, 0} = 0.6911, \Omega_{B, 0} = 0.0486, 
H_{0} = 67.74$ km s$^{-1}$ Mpc$^{-1}$ \citep[][results XIII]{Planck15}. 

The size of the cubic cosmological volume is $(2 h^{-1} ~ {\rm Mpc})^3$ comoving. 
Note that we adopt this volume in order to allow for the formation of a larger number 
of less massive (dwarf) galaxy systems than in typical larger box simulations. 
We used a single random realization of this box volume, and varied the 
sub-resolution physical parameters in the different simulations as described next. 
We use $256^3$ dark matter and $256^3$ gas particles in the initial condition. 
The dark matter particle mass is $m_{\rm DM} = 3.44 \times 10^{4} h^{-1} M_{\odot}$, 
and the gas particle mass is $m_{\rm gas} = 6.43 \times 10^{3} h^{-1} M_{\odot}$. 
The gravitational softening length is set as $L_{\rm soft} = 0.1 h^{-1}$ kpc comoving. 
Starting from $z = 100$, the box is subsequently evolved up to $z = 4$, 
with periodic boundary conditions. 





We execute a series of simulations, with characteristics listed in Table~\ref{Table-Sims}. 
All the runs incorporate metal cooling, chemical enrichment, SF and SN feedback. 
The first run has no BH, while the remaining runs include BHs. 
We vary 3 parameters (which were introduced in \S\ref{sec-num-BH}) 
exploring different BH sub-resolution models: 
the minimum halo mass for BH seeding $(M_{\rm HaloMin})$, the seed BH mass 
$(M_{\rm BHseed})$, and the kick velocity for BH kinetic feedback $(v_w)$. 
These parameters were observed to have the largest effect on the BH growth and feedback. 
The simulation names are formatted like below, 
where $X$, $Y$, $Z$ and $A$ are numbers in the BH runs. 

\begin{itemize} 

\item {\it SN} : no BH present. This is a control simulation in which 
only cooling, enrichment, star-formation, and SN feedback are implemented. 

\item {\it BHsXhYeZvA} : with BH accretion and feedback. 
The parameter values are, $M_{\rm BHseed} = 10^{X} M_{\odot}, 
M_{\rm HaloMin} = Y \times 10^{Z} M_{\odot}$, and $v_w = A \times 1000$ km/s. 

\end{itemize} 

Halos are first identified using the FOF algorithm. 
A halo is required to contain a minimum of $32$ particles. 
Galaxies are tracked within our simulations using the subhalo finder {\it SubFind}, 
which associates substructures to FOF halos. 
The minimum number of particles to make a valid subhalo is $20$. 
The centre of each galaxy is considered as the location of 
the gravitational potential minimum of its subhalo. 
We define galaxy stellar mass as the mass of all star particles inside 
the subhalos obtained by the subhalo finder {\it SubFind}. 
The halo mass $(M_{\rm halo})$ of a galaxy, and its virial radius in comoving 
coordinates $(R_{200})$, are related such that $R_{200}$ encloses a density 
$200$ times the mean comoving matter density of the Universe: 
\begin{equation} 
\label{eq-Mhalo} 
M_{\rm halo} = \frac{4 \pi}{3} R_{200}^3 \left(200 \rho_{\rm crit} \Omega_{M,0}\right) , 
\end{equation} 
where $\rho_{\rm crit} = 3 H_0^2 / (8 \pi G)$ is the present critical density.

\section{Results and Discussion} 
\label{sec-results}

\subsection{Large Scale Environment} 
\label{sec-res-LargeScale}

\begin{figure*} 
\centering 
\includegraphics[width = 1.0 \linewidth]{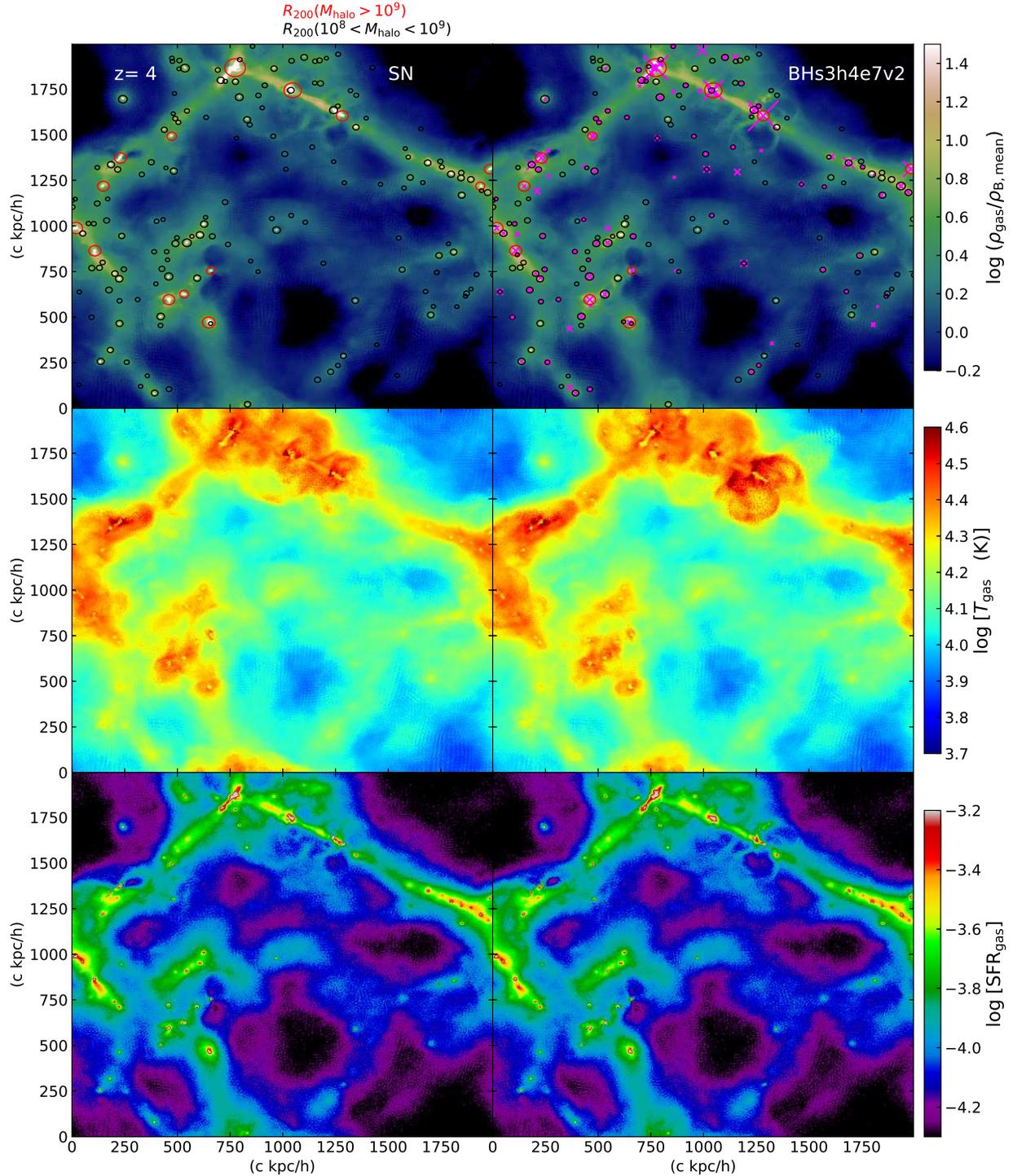} 
\vspace{-2.4cm} \\ 
\caption{ 
Gas properties (2D projection integrated over the third Cartesian direction) 
in the whole $(2000 h^{-1}$ kpc$)^3$ simulation box at $z = 4$, 
with the three rows presenting: 
overdensity (first row -- top), temperature (second), and SFR (third -- bottom). 
The two columns are for different simulations: 
{\it SN} (left) and {\it BHs3h4e7v2} (right). 
The red circles in the top row depict the virial radius $R_{\rm 200}$ of galaxies 
in the mass range $M_{\rm halo} > 10^{9} M_{\odot}$, while the black circles 
show the $R_{\rm 200}$ of $10^{8} < M_{\rm halo} < 10^{9} M_{\odot}$ galaxies. 
The positions of BHs in run {\it BHs3h4e7v2} are indicated by the 
magenta cross-points in the top-right panel, 
where the symbol size is proportional to BH mass. 
} 
\label{fig-Gas-rho-T-SFR} 
\end{figure*} 

The gas morphology in our simulation box, or the large scale structures, 
is plotted in Fig.~\ref{fig-Gas-rho-T-SFR}. 
It displays gas properties in the whole $(2000 h^{-1}$ kpc$)^3$ 
comoving volume at $z = 4$, for two simulations: 
{\it SN} (left column), and {\it BHs3h4e7v2} (right column) - 
which is a run that produced one of the most-massive BH at $z = 4$. 
The overdensity (i.e., the ratio between the gas density and the cosmological mean 
baryon density in the Universe), temperature, and star-formation rate (SFR) 
of the gas are plotted in the three rows from the top. 
The black circles in the top row depict the virial radius $R_{\rm 200}$ 
(defined in Eq.~\ref{eq-Mhalo}) of dwarf galaxies with halo masses in the range 
$10^{8} \leq M_{\rm halo} \leq 10^{9} M_{\odot}$. 
The red circles show the $R_{\rm 200}$ of relatively massive galaxies, those 
having higher halo masses $M_{\rm halo} > 10^{9} M_{\odot}$. 

The spatial locations of the BHs within our {\it BHs3h4e7v2} simulation box 
can be visualized in the top-right panel of Fig.~\ref{fig-Gas-rho-T-SFR}, 
overplotted with the gas overdensity. Here the magenta cross-points 
designate BH positions, with the symbol size proportional to BH mass. 
In this run, BHs are seeded at the centres of galaxies with 
$M_{\rm HaloMin} = 4 \times 10^{7} M_{\odot}$. 
Therefore all the red circles ($M_{\rm halo} > 10^{9} M_{\odot}$ galaxies) 
and most of the black circles ($M_{\rm halo} = 10^{8}-10^{9} M_{\odot}$ 
galaxies) contain BHs at their centres. 

The cosmological large-scale-structure filaments are visible 
in all the panels of both the runs. 
There are three Mpc-scale filaments: 
extending from east to north, from west to north, and from west to south. 
In addition, there is an overdense region running 
from the center of the box to the south-west. 
The filaments consist of dense (yellow and white regions in the top panels), and 
star-forming (lightgreen, yellow, red and white regions in the bottom panels) gas. 
The massive galaxies (red circles) lie at the high-density intersections 
of the filaments, or in the filaments. 

In terms of temperature, the immediate vicinity of the dense filaments consists of 
hotter gas ($T \sim 10^{4.6}$ K, red regions in the middle panels of 
Fig.~\ref{fig-Gas-rho-T-SFR}), as compared to that in the low-density 
intergalactic medium and voids (yellow and blue regions). 
Several mechanisms play together to heat the gas to higher temperatures 
in the filament vicinity. 
There are global environmental processes like shock heating during 
galaxy mergers, and large-scale-structure formation, 
which are present in both the {\it SN} and {\it BHs3h4e7v2} runs. 
Acting together there are local galactic processes like 
feedback driven by SN (present in both the columns), 
and BHs (present in the right column only), which heats the gas, 
and also often generate outflows. 

As they accrete and grow, the BHs provide feedback energy 
(according to the prescription described in \S\ref{sec-num-BH}), 
which may drive gas outflows. 
High-velocity gas ejected by central BH feedback propagates radially outward, 
and shocks with the surrounding slower-moving gas, creating bubble-like gas outflows. 
Our simulation {\it BHs3h4e7v2} shows the formation of BH feedback-induced outflows, 
as can be seen in Fig.~\ref{fig-Gas-rho-T-SFR} in the top-right half 
of the right column panels, around the most-massive BH. 
The outflows are extended bipolar oval-shaped regions along 
the north-east to south-west direction, propagating to about $10 \times R_{\rm 200}$. 
The outflows consist of hot ($T > 10^{4.6}$ K) - visible as red areas 
in the temperature map (middle-right panel), and low-density gas (top-right panel).

\subsection{Black Hole Accretion and Growth} 
\label{sec-res-BH-Growth} 

\begin{figure*} 
\centering 
\includegraphics[width = 0.8 \linewidth]{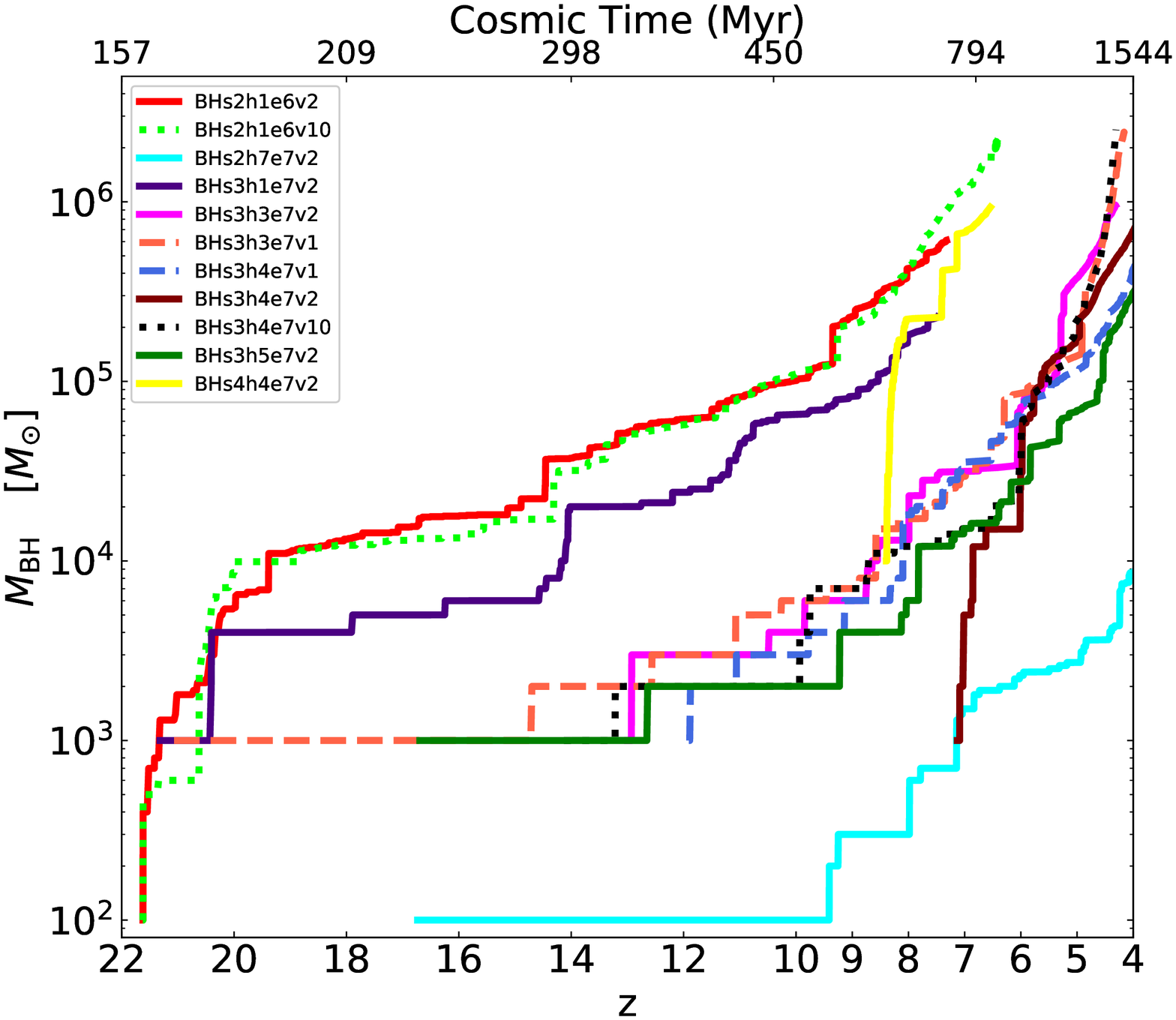} 
\caption{ 
BH mass growth with redshift of the most-massive BH in each run. 
The different colours and linestyles discriminate the runs as labelled, which are: 
{\it BHs2h1e6v2} - red solid curve, 
{\it BHs2h1e6v10} - lime-green dotted, 
{\it BHs2h7e7v2} - cyan solid, 
{\it BHs3h1e7v2} - indigo solid, 
{\it BHs3h3e7v2} - magenta solid, 
{\it BHs3h3e7v1} - light-red dashed, 
{\it BHs3h4e7v1} - blue dashed, 
{\it BHs3h4e7v2} - maroon solid, 
{\it BHs3h4e7v10} - black dotted, 
{\it BHs3h5e7v2} - green solid, and 
{\it BHs4h4e7v2} - yellow solid. 
} 
\label{fig-BH-Mass-vs-Time} 
\end{figure*} 

Fig.~\ref{fig-BH-Mass-vs-Time} presents the BH mass growth with redshift 
of the most-massive BH for eleven simulations of Table~\ref{Table-Sims}. 
The first BHs are seeded at $z \sim 18 - 25$ in our simulations, when the first 
halos reach the corresponding lower limit 
$M_{\rm HaloMin} = (10^{6} - 7 \times 10^{7}) M_{\odot}$. 
In some of the runs, one of these first seeds grow to become the most-massive BH. 
However in runs {\it BHs3h4e7v2} (maroon solid curve) and 
{\it BHs4h4e7v2} (yellow solid curve), 
the BH which becomes most-massive is seeded at later epochs $z \sim 7.5 - 12$. 
This variance in the seed epochs is because of the different BH growth modes, 
as described next. 

Each BH starts as an initial seed of $M_{\rm BHseed} = (10^{2}, 10^{3}, 10^{4}) M_{\odot}$. 
The subsequent growth is due to merger with other BHs and gas accretion. 
When two galaxies containing BHs merge during hierarchical structure formation, 
their central BHs merge as well (according to the prescription in 
\S\ref{sec-num-BH}) to form a single larger BH. 
In addition, BHs grow in mass by accreting dense gas from their surroundings, 
as galaxies evolve and gas inflows to their centres. 

We find that in general, when seeded in larger halos, BHs start to grow later and 
have an overall smaller growth at the same redshift 
(comparing indigo and magenta curves, for instance). 
Furthermore, larger seed BHs grow more massive earlier than smaller seeds 
(comparing yellow and maroon curves), as naively expected. 
We also find that varying the $M_{\rm HaloMin}$ and $M_{\rm BHseed}$ together 
has a competing effect on the BH mass growth, which might cancel each other. 
A BH can grow similarly when seeded in smaller halos with a smaller seed mass 
(red curve), as in larger halos with a higher seed mass (indigo curve). 

Comparing the blue-dashed (run {\it BHs3h4e7v1}) versus maroon-solid (run {\it BHs3h4e7v2}) 
curves, we find that the final BH masses at $z = 4$ are comparable. 
However there is a significant difference in the epoch at which the BHs are seeded, 
while the only parameter variation between these runs is the BH wind velocity. 
This occurs because Fig.~\ref{fig-BH-Mass-vs-Time} shows only the most-massive BH, 
and there are variations in the growth assembly of BHs caused by a coupling 
between the underlying physics and cosmological hierarchical mergers. 
In one simulation there can be $2$ massive BHs of comparable masses 
(e.g., seen by the magenta cross-points in Fig.~\ref{fig-Gas-rho-T-SFR} 
top-right panel), which were seeded at different epochs. 
So even though the seeding in these runs should be uniform, 
in {\it BHs3h4e7v2} a BH which was seeded at a later epoch $z=7$ grows 
to become the largest, whereas in {\it BHs3h4e7v1} the largest was seeded at $z=12$.

\begin{figure*} 
\centering 
\includegraphics[width = 1.15 \linewidth]{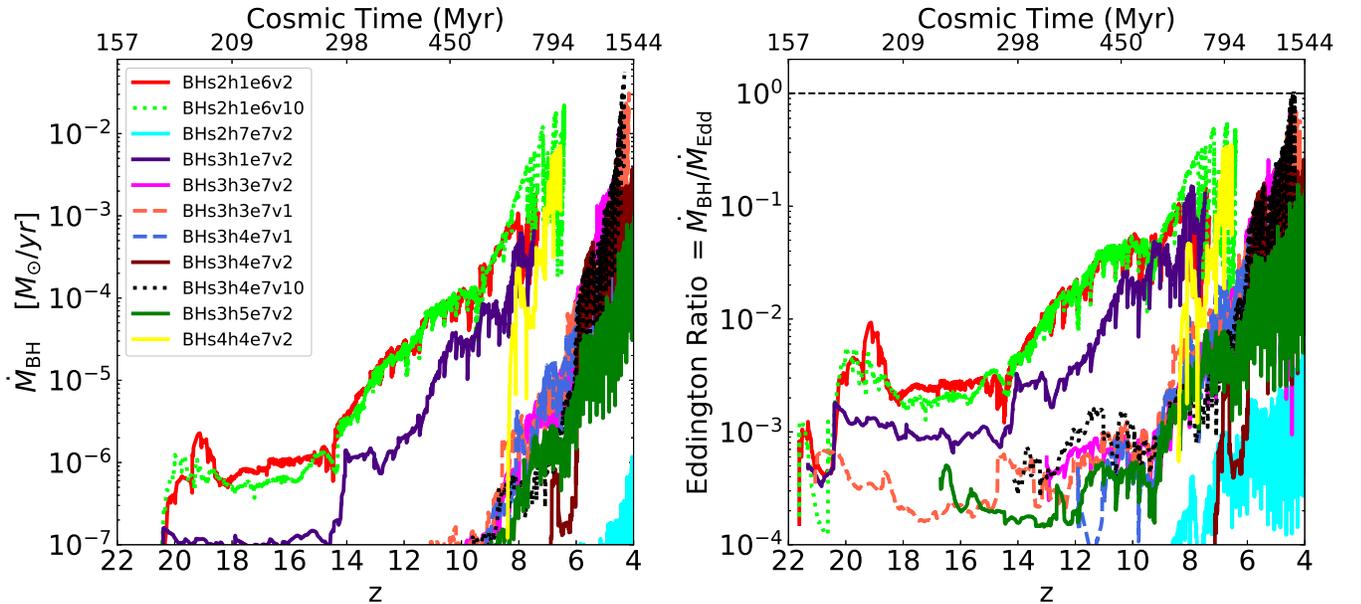} 
\caption{ 
BH accretion rate (left panel) and Eddington ratio (right panel) growth with redshift 
of the most-massive BH in each run. 
} 
\label{fig-BH-AccRate-vs-Time} 
\end{figure*} 

Fig.~\ref{fig-BH-AccRate-vs-Time} displays the accretion rate evolution 
with redshift, of the most-massive BH in each run: 
BH mass accretion rate ($\dot{M}_{\rm BH}$) in the left panel, 
and Eddington ratio $= \dot{M}_{\rm BH} / \dot{M}_{\rm Edd}$ at the right. 
There is variability of the $\dot{M}_{\rm BH}$, 
whereby it fluctuates by a factor of up to $100$. 
In the beginning, when the BHs are just seeded, they are accreting at highly 
sub-Eddington rates with Eddington ratio $< 0.0001 - 0.001$. 
The accretion rate grows with time as the BHs grow in mass, and the gas density 
in the BH vicinity increases by cosmic gas inflow to galaxy centers. 

We note that the growth of a BH is initially dominated by mergers with other BHs 
soon after its seeding for about a $100$ to $200$ Myr, 
when $\dot{M}_{\rm BH} <\sim 0.05 \dot{M}_{\rm Edd}$. 
Growth by gas accretion dominates at later times, when the Eddington ratio is larger. 
Such a trend was also found for SMBH growth in \citet{Barai18}. 
In some of our simulations, the BHs continue to grow at much later times, 
even when star formation has started to be quenched. 
This occurrs due to gas inflows driven by galaxy mergers and galaxy tidal encounters 
(further discussed in \S\ref{sec-res-Vary-SN-strength-BH-growth}). 

We find that the massive BHs grow to $M_{\rm BH} =$ a few$\times 10^{5} - 10^{6} M_{\odot}$ 
by $z = 5$ in most of our simulations 
(red solid, lime-green dotted, indigo solid, magenta solid, light-red dashed, 
blue dashed, maroon solid, black dotted, green solid, and yellow solid curves). 
The accretion rate of these massive BHs has grown to 
$\dot{M}_{\rm BH} = (0.2 - 1) \dot{M}_{\rm Edd}$. 
The exception is the case where BHs of the smallest seed $(10^{2} M_{\odot})$ are placed 
in the largest halos $(7 \times 10^{7} M_{\odot})$ (run {\it BHs2h7e7v2}, cyan solid curve), 
where the most-massive BH grows up to $M_{\rm BH} = 10^4 M_{\odot}$ only. 
For this BH, gas accretion is always occurring at low Eddington ratios 
$(\dot{M}_{\rm BH} \leq 0.01 \dot{M}_{\rm Edd})$. 

The final BH properties reached at $z = 4$ depend on the simulation. 
E.g., in run {\it BHs3h3e7v1} (light-red dashed curve): 
$M_{\rm BH} = 2 \times 10^6 M_{\odot}$ and $\dot{M}_{\rm BH} = 0.02 M_{\odot}$/yr. 
In Table~\ref{Table-Final-BH-mass} we list the final BH mass and the 
accretion rate range (maximum and minimum over the final $50$ Myr period) 
of the most-massive BH, together with the final redshift of each simulation run. 

We note that some simulations were stopped before $z = 4$. 
These are the runs where star-formation is already quenched at a high-$z$. 
To optimize our finite computational resources, once significant quenching 
of SF were already achieved by the growing BHs at a higher-$z$, 
we did not continue those simulations further.

%

\begin{table*} 
\begin{minipage}{1.0 \linewidth} 
\caption{ 
Final black hole mass and accretion rate range of the most-massive BH for the different simulation runs. 
} 
\label{Table-Final-BH-mass} 
\centering 
\begin{tabular}{@{}ccccc} 

\hline 

Run  & Final     & BH mass,                 & BH accretion rate, $\dot{M}_{\rm BH} [M_{\odot}/yr]$ \\ 
name & Redshift & $M_{\rm BH} [M_{\odot}]$  & (maximum and minimum over the final $50$ Myr period) \\ 

\hline 

{\it BHs2h1e6v2}  & $7.5$ & $6 \times 10^{5}$   & $0.001 - 0.002$ \\ 

{\it BHs2h1e6v10} & $6.5$ & $3 \times 10^{6}$   & $0.0002 - 0.03$ \\ 

{\it BHs2h4e7v2}  & $4.2$ & $2 \times 10^{4}$   & $6 \times 10^{-7} - 10^{-5}$ \\ 

{\it BHs2h7e7v2}  & $4$   & $1 \times 10^{4}$   & $10^{-7} - 10^{-6}$ \\ 

{\it BHs3h1e7v2}  & $7.5$ & $3 \times 10^{5}$   & $0.0002 - 0.0007$ \\ 

{\it BHs3h2e7v2}  & $6.2$ & $3.5 \times 10^{5}$ & $0.0007 - 0.001$ \\ 

{\it BHs3h3e7v2}  & $4.2$ & $8 \times 10^{5}$   & $0.003 - 0.015$ \\ 

{\it BHs3h3e7v1}  & $4.2$ & $3 \times 10^{6}$   & $0.002 - 0.03$ \\ 

{\it BHs3h4e7v1}  & $4$   & $4 \times 10^{5}$   & $0.0005 - 0.002$ \\ 

{\it BHs3h4e7v2}  & $4$   & $4 \times 10^{5}$   & $0.0002 - 0.004$ \\ 

{\it BHs3h4e7v5}  & $4$   & $2 \times 10^{6}$   & $0.005 - 0.03$ \\ 

{\it BHs3h4e7v10} & $4.2$ & $3 \times 10^{6}$   & $0.003 - 0.06$ \\ 

{\it BHs3h5e7v2}  & $4$   & $3 \times 10^{5}$   & $6 \times 10^{-5} - 0.0008$ \\ 

{\it BHs4h4e7v2}  & $6.5$ & $1 \times 10^{6}$   & $0.0015 - 0.008$ \\ 

\hline 
\end{tabular} 

\end{minipage} 
\end{table*} 


The left panel of Fig.~\ref{fig-SFRD} shows the total BH Accretion Rate Density 
(BHARD) as a function of redshift. 
Each curve is for one run, labelled by the different colours and linestyles. 
The BHARD (in $M_{\odot}$ yr$^{-1}$ Mpc$^{-3}$) is computed by summing over 
the accretion rate of all the BHs inside each simulation box at a time,
and dividing it by the box volume. 
We find an overall similar behaviour of the most-massive BH accretion rate 
(Fig.~\ref{fig-BH-AccRate-vs-Time}), 
and the total BHARD in the simulation volume (Fig.~\ref{fig-SFRD}). 
This is because the relative accretion rate trends between simulations 
remain the same when considering one BH or multiple BHs. 
In other words, all the BHs in a simulation behave similarly in this regard. 
E.g. if the $\dot{M}_{\rm BH}$ of the most-massive BH in a run 
is higher than the most-massive $\dot{M}_{\rm BH}$ of another run, 
then the BHARD is also higher in the former case, because it will have 
several BHs whose $\dot{M}_{\rm BH}$ are higher than in the 2nd run.

\subsection{Star Formation Rate} 
\label{sec-res-SFRD} 

\begin{figure*} 
\centering 
$ 
\begin{array}{c} 
\includegraphics[width = 0.52 \linewidth]{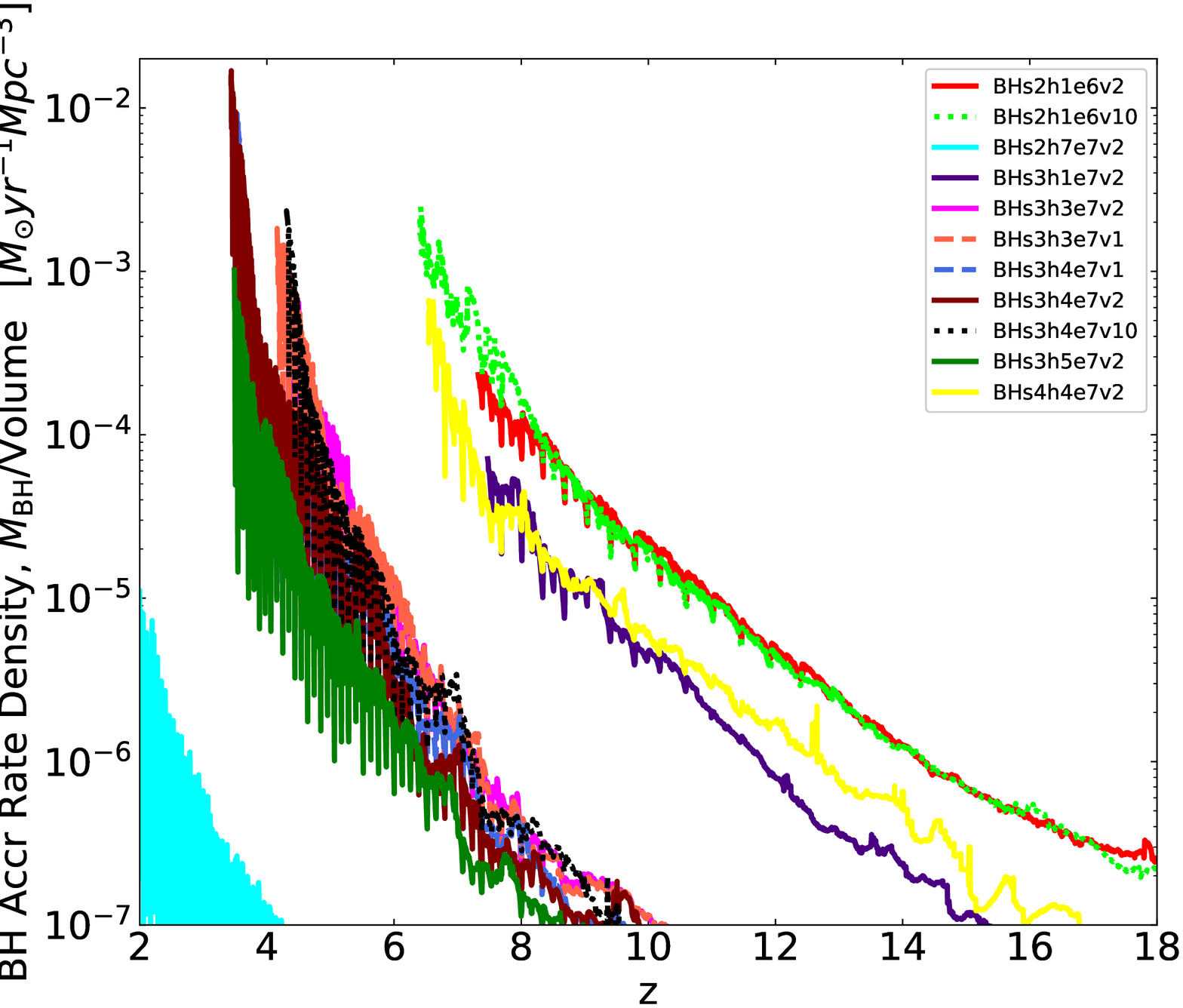} 
\includegraphics[width = 0.52 \linewidth]{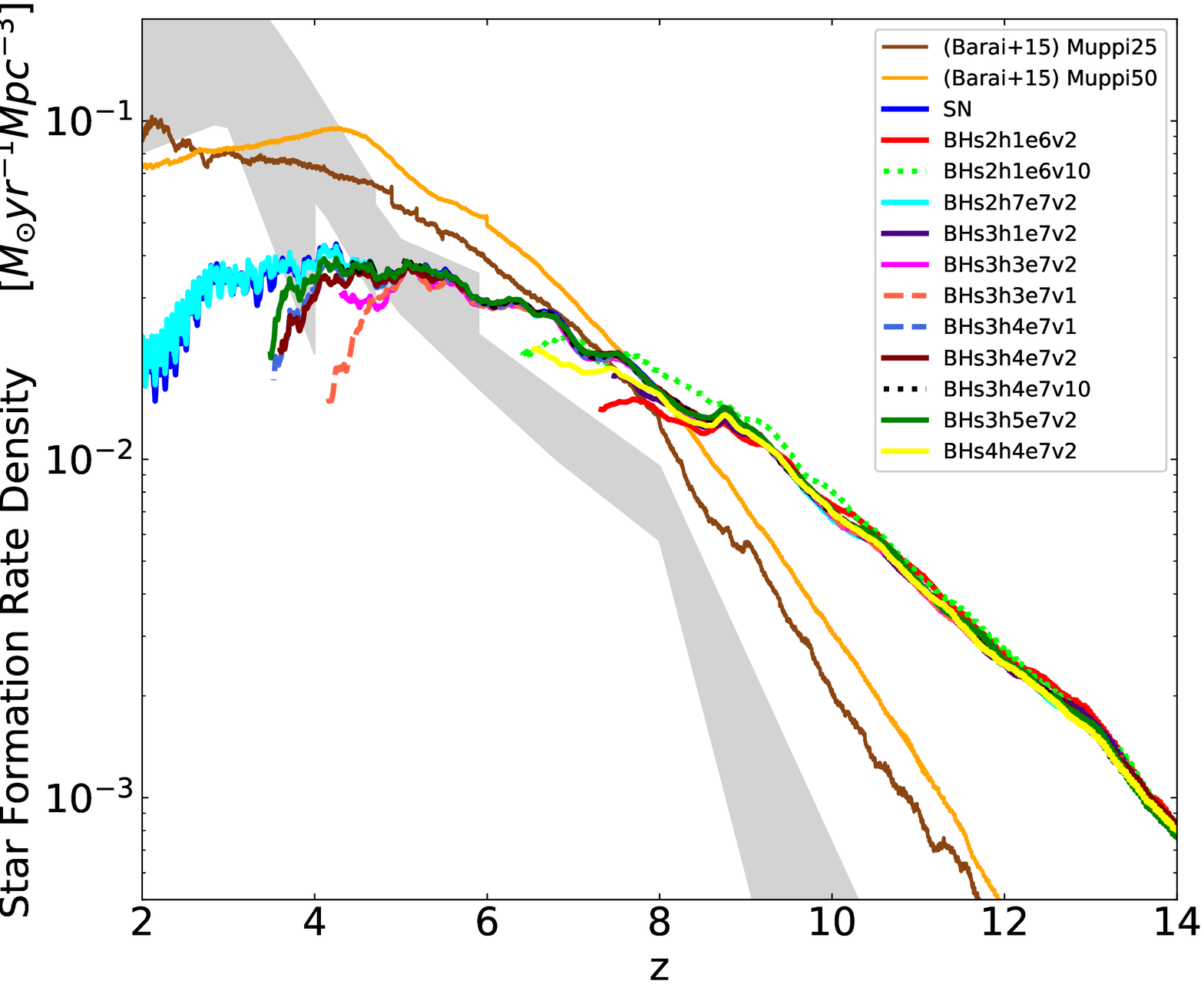} 
\end{array} 
$ 
\caption{ 
Total BH accretion rate density (left panel), and star formation rate density 
(right panel) integrated over the whole simulation volume as a function of 
redshift, with the different models labelled by the colours and linestyles. 
For comparison in the right panel, the brown and orange thinner curves show 
two models from \citet[][]{Barai15}, which are SFRD results for more massive 
galaxy formation/evolution, from larger volume cosmological simulations. 
The grey shaded region denotes a combination of observational SFRD data range from 
\citet{Cucciati12}, and the compilations therein originally from 
\citet{PerezGonzalez05}, \citet{Schiminovich05}, \citet{Bouwens09}, 
\citet{Reddy09}, \citet{Rodighiero10}, \citet{vanderBurg10}, \citet{Bouwens12}. 
} 
\label{fig-SFRD} 
\end{figure*} 

\begin{figure*} 
\centering 
\includegraphics[width = 1.15 \linewidth]{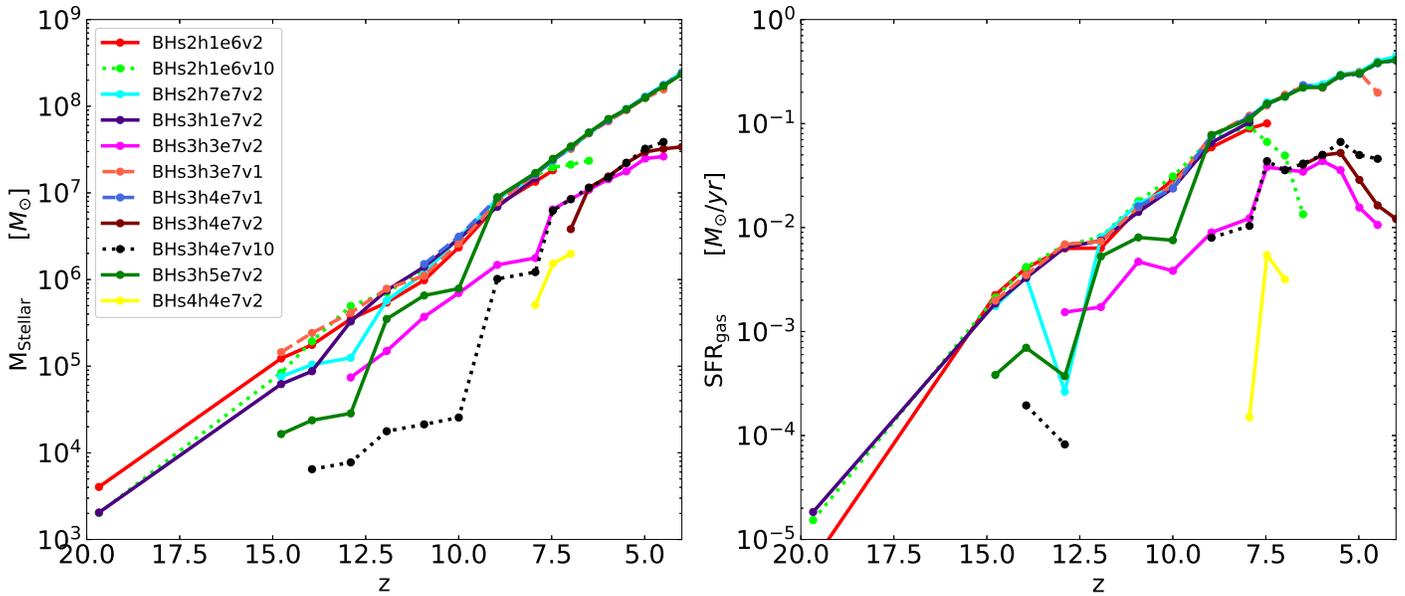} 
\caption{ 
Stellar mass (left panel) and total star formation rate (right panel) 
evolution with redshift of the galaxy hosting the most-massive BH in each run. 
} 
\label{fig-StellarMass-and-SFR} 
\end{figure*} 

Stars form in the simulation volume from cold dense gas. 
Fig.~\ref{fig-SFRD} exhibits the global Star Formation Rate Density (SFRD) as a function 
of redshift, for all the simulations labelled by the different colours and linestyles. 
The SFRD (in $M_{\odot}$ yr$^{-1}$ Mpc$^{-3}$) is computed by summing over 
all the SF occurring in each simulation box at a time, 
and dividing it by the time-step interval and the box volume. 
Observational data ranges are shown, for a comparison, as the grey shaded region, 
taken mainly from \citet{Cucciati12}, and the compilations therein originally from 
\citet{PerezGonzalez05}, \citet{Schiminovich05}, \citet{Bouwens09}, 
\citet{Reddy09}, \citet{Rodighiero10}, \citet{vanderBurg10}, \citet{Bouwens12}. 
We note that the observational data are mostly based on massive galaxies. 
The SFRD as indicated by these observations continues to grow at $z < 6$, and has 
a peak around $z = 2$ which is due to the brightest galaxies \citep{Cucciati12}. 
This implies that signiﬁcant gas reservoirs exist in such more massive galaxies 
at these epochs, and are being replenished by cosmic accretion and galaxy mergers. 
While quenching processes due to BH and SN driven outflows/feedback 
are still less strong to lower the SF in these galaxies at $z \sim 2 - 3$. 


In our simulations of the formation of dwarf galaxies, we see that 
the SFRD rises with time from early epochs $z \sim 15$, 
and all the runs behave similarly up to $z \sim 9$. 
Most of the simulations reach a maximum SFRD in the form of a plateau between 
$z = 4 - 6$, and the SFRD decreases at $z > 6$ and at $z < 4$. 
The scenario changes completely at $z < 8$ if we consider no BH or SN feedback 
(see Fig.~\ref{fig-SFRDvarySN} and \S\ref{sec-res-Vary-SN-strength-SFR}). 
In the absence of any feedback, the SFRD continues to grow until recent epochs 
overshooting the observational limits 
(more details in \S\ref{sec-res-Vary-SN-strength-SFR}). 

Back to  Fig.~\ref{fig-SFRD}, we consider the {\it SN} run (dark blue curve) 
without BHs as the baseline, 
and compare other simulations with it to estimate the impact of BH feedback. 
The SFRD in simulation {\it BHs2h7e7v2} (cyan solid curve) is almost similar to that in 
the run {\it SN}, because the BHs are too small there to generate enough feedback. 
A similar outcome happens in the other runs at $z \geq 9$, when the BHs are too small. 

Star formation mostly occurs over an extended region at galaxy centres, 
where cosmic large-scale-structure gas inflows and cools. 
The presence of a central BH helps to quench star formation, because a fraction of 
dense gas is accreted onto the BH, and a fraction is ejected out by BH feedback. 
The processes of BH accretion and feedback contribute to suppress SF substantially 
in most of the runs, from $z \sim 9$ onwards, 
when the BHs have grown massive and generate larger feedback energy. 
Compared to the {\it SN} case (dark blue curve), 
the SFRD is reduced by a factor of several in most of the other runs at $z < 9$. 

We find that BH accretion and feedback causes a quenching of the SFRD 
at cosmic epochs $z \leq 9$, by factors $1.1 - 3$ times reduction. 
The precise redshifts and suppression factors are given in the following 
for the different simulations (with the same BH outflow velocity $v_w = 2000$ km/s): 
\begin{itemize} 
\item {\it BHs2h1e6v2} (red solid curve in Fig.~\ref{fig-SFRD}) 
at $z \leq 9$ up to $2$ times reduction of SFRD, 
\item {\it BHs3h1e7v2} (indigo solid curve) at $z \leq 8$ up to $1.1$ times, 
\item {\it BHs4h4e7v2} (yellow solid curve) at $z \leq 8$ up to $2$ times, 
\item {\it BHs3h2e7v2} at $z \leq 7$ up to $1.5$ times, 
\item {\it BHs3h3e7v2} (magenta solid curve) at $z \leq 5$ up to $2$ times, 
\item {\it BHs3h4e7v2} (maroon solid curve) at $z \leq 4.5$ up to $3$ times, 
\item {\it BHs3h5e7v2} (green solid curve) at $z \leq 4.2$ up to $3$ times. 
\end{itemize} 
Thus, we find that BHs need to grow to a mass $M_{\rm BH} \sim$ a few times 
$10^5 M_{\odot}$, in order to suppress star-formation in dwarf galaxies 
(of $M_{\star} = 10^{5} - 10^{8} M_{\odot}$). 

A larger BH outflow velocity ($v_w$) does not change the results substantially. E.g., 
comparing the red solid and lime-green dotted curves in Fig.~\ref{fig-BH-Mass-vs-Time}, 
the BH mass growth remains almost the same with $v_w = 2000$ and $10000$ km/s. 
Increasing $v_w$ decreases the mass outflow rate 
($\dot{M}_w$ in Eq.(\ref{eq-MdotW-EDW})), 
in the way favoring an increase in the BH accretion rate, 
as we see in Fig.~\ref{fig-BH-AccRate-vs-Time} 
at late epochs (comparing red solid and lime-green dotted curves at $z < 9$). 
Further, the smaller mass outflow rate to the environment delays the suppression 
of star formation (comparing lime-green dotted and red solid curves in Fig.~\ref{fig-SFRD}). 



Fig.~\ref{fig-StellarMass-and-SFR} displays the stellar properties as a 
function of redshift of the galaxy hosting the most-massive BH in each run 
(labelled by its colour and linestyle): 
stellar mass ($M_{\star}$ in units of $M_{\odot}$) in the left panel, 
and total star formation rate in the right panel. 
The total SFR (in $M_{\odot}/yr$) is computed by summing over 
all the SF in gas particles occurring inside the galaxy virial radius $R_{\rm 200}$ 
(defined in Eq.~\ref{eq-Mhalo}). 

We see that the SFR of the galaxy hosting the most-massive BH behaves 
similarly overall as the global SFRD (presented in Fig.~\ref{fig-SFRD} - right panel). 
The SFR continues to increase (with no quenching) in the simulation {\it BHs2h7e7v2} 
(cyan solid curve), because there the BHs are too small to generate enough feedback. 
A similar non-quenching of SFR is also seen in run {\it BHs3h5e7v2} (green solid curve), 
where the BH mass at $z = 4$ is smaller than all the runs. 
For the other simulations, there is a quenching of SFR in the galaxies 
by their centrally growing BHs. This can be visualised by the anti-correlation: 
SFR in Fig.~\ref{fig-StellarMass-and-SFR} reduces 
when the BH mass has a corresponding sharp growth in Fig.~\ref{fig-BH-Mass-vs-Time}. 

\begin{figure*} 
\centering 
\includegraphics[width = 1.12 \linewidth]{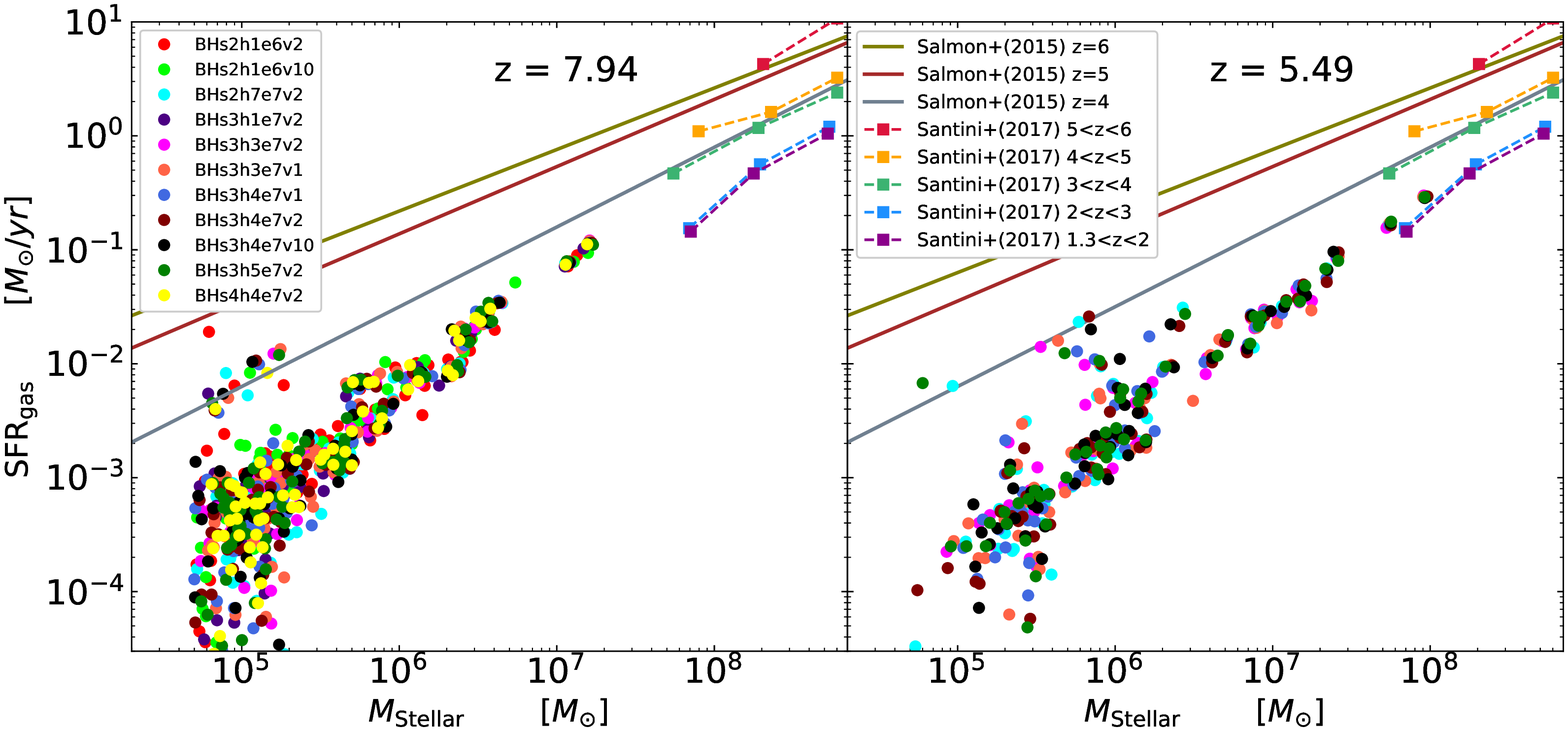} 
\caption{ 
SFR versus stellar mass of the galaxies within the simulation volume, 
at the epoch $z = 7.94$ in the left, and $z = 5.49$ in the right panel. 
The filled circles are our simulation results, with the plotting colours 
distinguishing results from different runs. 
The lines indicate observational results of SFR versus stellar mass relation for: 
star-forming galaxies at $z = 4, 5, 6$ \citep{Salmon15} as the solid lines, and 
galaxies in different redshift bins within $z = 1.3 - 6$ \citep{Santini17} 
as the dashed lines with square plotting symbols. 
} 
\label{fig-SFR-vs-StellarMass} 
\end{figure*} 

Fig.~\ref{fig-SFR-vs-StellarMass} presents the 
SFR (in $M_{\odot}/yr$) versus stellar mass ($M_{\star}$) of all the galaxies within the 
cosmological boxes, at the epoch $z = 7.94$ in the left panel, and $z = 5.49$ in the right. 
The filled circles are our simulation results, with the plotting colours 
distinguishing results from different runs. 
Observational data is overplotted as the solid and dashed lines 
indicating the SFR versus $M_{\star}$ relationships at different epochs. 
The solid lines indicate observations from \citet{Salmon15} 
of star-forming galaxies at $z = 4, 5, 6$: the best-fit relation of SFR versus 
$M_{\star}$ as written in their Equation (6) with parameters given in their Table 4. 
Note that the fit was done on more massive galaxies, 
but here the same relation has been extended to lower masses. 
The dashed lines show observations from \citet{Santini17} 
in different redshift bins within $z = 1.3 - 6$: the data points are taken from their 
Table 1, and here only the average data with $M_{\star} < 10^{9} M_{\odot}$ 
are shown as the square plotting symbols. 

We find that our simulated DGs exhibit a positive SFR versus $M_{\star}$ 
correlation qualitatively similar to the observations, but with an offset 
in the $[$SFR $- M_{\star}]$ plane. 
Most of our DGs are consistent with the relations showing the observational 
main sequence of star-forming galaxies, when extrapolated to lower masses. 
There is a large scatter in our SFR versus $M_{\star}$ correlation, 
going up to $1$ dex at the lowest masses. 
Some of our low-mass ($M_{\star} \sim 10^{5} - 10^{6} M_{\odot}$) and 
high-SFR (SFR $\sim 0.003 - 0.03 M_{\odot}/yr$) galaxies fall on the 
$z \sim 4$ relation of \citet{Salmon15} and \citet{Santini17}. 
At $z = 5.49$ (right panel of Fig.~\ref{fig-SFR-vs-StellarMass}), 
our galaxies lie on the $z \sim 2$ relation of \citet{Santini17}. 
Whereas at $z = 7.94$ (left panel), our galaxies 
lie between the $z \sim 2$ and $z \sim 4$ relations of \citet{Santini17}.

\subsection{Black Hole - Galaxy Correlation} 
\label{sec-res-BH-Gal-CoEvol} 

\begin{figure*} 
\centering 
\includegraphics[width = 1.12 \linewidth]{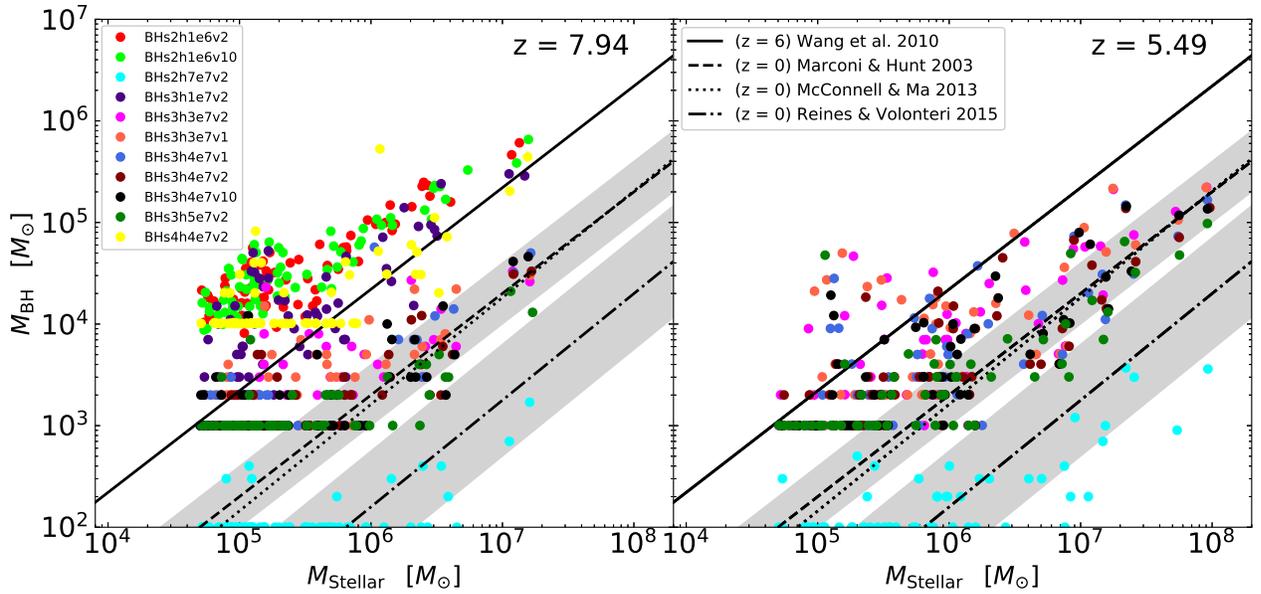} 
\caption{ 
BH mass versus stellar mass of the galaxies within the simulation volume, 
at two epochs $z = 7.94, 5.49$, in the two panels. 
The plotting colours distinguish results from different runs. 
The black lines indicate the observed BH mass versus galaxy stellar mass 
relation for: $z \sim 6$ quasars \citep{Wang10} as the solid line, and 
local galaxies \citep{Reines15} as the dash-dotted line, 
showing the correlation with the bulge mass as the dashed \citep{Marconi03} 
and dotted \citep{McConnell13} lines. 
} 
\label{fig-Mass-BH-vs-Stellar} 
\end{figure*} 

The BH - galaxy correlation obtained in our simulations is presented 
in Fig.~\ref{fig-Mass-BH-vs-Stellar} as the $M_{\rm BH}$ 
versus $M_{\star}$ (stellar mass) diagram. It shows all the galaxies 
within the simulation volume with $M_{\star} > 10^{4.5} M_{\odot}$, 
at two epochs: $z = 7.94$ in the left panel, and $z = 5.49$ in the right panel. 
The plotting colour distinguishes results from different runs. 
Observational data is overplotted as the black lines 
indicating the BH mass versus stellar bulge mass relationships at different epochs. 
Local galaxies ($z = 0$) are represented by the black-dashed line: 
$M_{\rm BH} / M_{\star} = 0.002$ \citep{Marconi03}, 
as well as from \citet{Reines15} as the dash-dotted line, 
and from \citet{McConnell13} as the dotted line. 
The ratio is observed to be steeper at high-$z$. 
Far-IR and CO bright $z \sim 6$ quasars lie along the black-solid line: 
median $M_{\rm BH} / M_{\star} = 0.022$ \citep{Wang10}. 

We find a huge scatter in the $[M_{\rm BH} - M_{\star}]$ correlation 
of the simulated galaxies, especially at low-$M_{\star}$. 
In our simulations {\it BHs2h1e6v2} (red symbols in Fig.~\ref{fig-Mass-BH-vs-Stellar}) 
and {\it BHs3h1e7v2} (indigo symbols), 
dwarf galaxies with stellar masses between $M_{\star} = 10^{4} - 10^{7} M_{\odot}$ 
contain BHs in the range $M_{\rm BH} = 10^{3} - 10^{5} M_{\odot}$ at $z = 7.9$, 
which are hence already more massive than BHs following the local relation, 
as well as BHs in $z \sim 6$ quasars. 
These two are also the runs where SF quenching happens the earliest, 
implying that the suppression occurs due to BH activities. 

BHs in the simulation {\it BHs4h4e7v2} (yellow symbols in 
Fig.~\ref{fig-Mass-BH-vs-Stellar}) fall on the observed $z \sim 6$ 
relation of $[M_{\rm BH} - M_{\star}]$ already at $z = 7.9$. 
In the run {\it BHs3h2e7v2}, BHs lie in between the two correlations at 
$z = 7.9$, and have migrated to the $z \sim 6$ relation at the later epoch. 
Central BHs in the runs {\it BHs3h3e7v2} (magenta symbols), 
{\it BHs3h4e7v2} (maroon symbols), and {\it BHs3h5e7v2} (green symbols) 
more or less follow the local $z = 0$ correlation at both the epochs plotted; 
although their scatter increases substantially in the right panel. 
When BHs do not grow much (run {\it BHs2h7e7v2}, cyan symbols, 
most-massive reaching $M_{\rm BH} < 10^4 M_{\odot}$ only), they always lie 
significantly below both of the $[M_{\rm BH} - M_{\star}]$ correlations. 

At the same time, there are studies 
\citep[e.g.,][using semi-analytical models]{Volonteri09}, 
which find that the existence of the $[M_{\rm BH} - \sigma]$ correlation 
is purely a reflection of the merging hierarchy of massive dark matter haloes. 

We find that there is a direct connection between early BH growth 
and the quenching of SF, which is henceforth caused by resulting BH feedback. 
In addition, our results of fast BH growth at the centers of dwarf galaxies 
suggest that these BHs grow faster than their host galaxies in the early Universe.


\subsection{Varying the SN Feedback} 
\label{sec-res-Vary-SN-strength} 

We now check the impact of supernovae feedback strength on our results, by varying the 
mass loading factor $\eta$ of SN wind kinetic feedback (\S\ref{sec-num-cool-SF-SN}). 
So far, we have discussed only simulations with the same SN feedback, 
using a default value of $\eta = 2$ \citep[e.g.,][]{Tescari11, Barai15}. 
This corresponds to $\chi = 0.27$ ($\chi$ defined in Eq.~(\ref{eq-SN-feedback}) 
of \S\ref{sec-num-cool-SF-SN}) of SN energy being carried away by the wind. 
We now consider a new run with a stronger SN-driven mass ejection. 
This simulation {\it BHs3h1e7v2-SNhi} has a higher $\eta = 5$, 
while the other parameters are the same as {\it BHs3h1e7v2}. 
In this new run, $\chi = 0.67$ of SN energy is carried away by the wind, 
which is a high value for the SN feedback efficiency.

\subsubsection{Impact of SN Feedback Strength on SFR} 
\label{sec-res-Vary-SN-strength-SFR} 

\begin{figure*} 
\centering 
\includegraphics[width = 0.6 \linewidth]{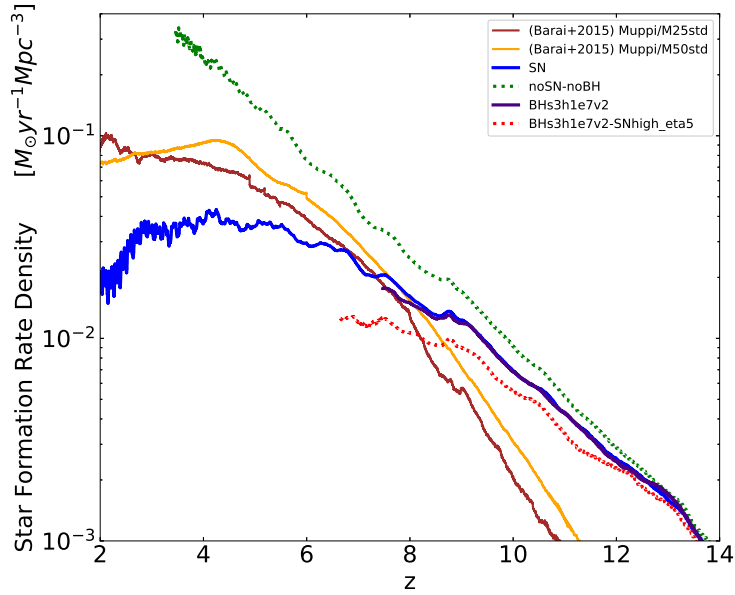} 
\caption{ 
Total star formation rate density as a function of redshift, 
in the simulations with different strengths of SN feedback. 
The colours and linestyles discriminate the runs as labelled, 
{\it noSN-noBH}       : green dashed, 
{\it SN}              : dark blue solid, 
{\it BHs3h1e7v2}      : indigo solid, and 
{\it BHs3h1e7v2-SNhi} : red dashed. 
For comparison, the brown and orange thinner curves show two models from 
\citet[][]{Barai15}, which are SFRD results for more massive galaxy 
formation/evolution, from larger volume cosmological simulations. 
} 
\label{fig-SFRDvarySN} 
\end{figure*} 

Fig.~\ref{fig-SFRDvarySN} displays the global Star Formation Rate Density as a 
function of redshift, for the simulations with different strengths of SN feedback. 
The SFRD rises with time as more and more stars form from available gas. 
The simulation {\it noSN-noBH} (green dashed curve) with no SN feedback and no BH 
accretion/feedback, has no source to deplete the gas, and the SFRD continues to 
rise in this run up to $z = 3$, overshooting the observational limits. 
Feedback processes from SN and BH suppress SF substantially 
in the other runs at $z < 10$. 

Comparing the cases {\it noSN-noBH} (green dashed) and {\it SN} (dark blue solid), 
we find that SN feedback quenches SF by $10 - 100$ times at $z < 6$. 
While the quenching by BH feedback (\S\ref{sec-res-SFRD}) is to 
reduce the SFRD by factors $1.1 - 3$ only. 
When the SN feedback strength is increased to $\eta = 5$ 
(simulation {\it BHs3h1e7v2-SNhi}, red dashed curve), the SFRD is lowered by a 
factor $2$ from $z = 10$ compared to run {\it BHs3h1e7v2} (indigo solid). 

Thus, we find that SN feedback quenches star-formation 
by a larger factor than the quenching by BH feedback. 
This result of ours is in agreement with several other galaxy simulation studies 
\citep[e.g.,][]{Dubois15, Habouzit17, Angles-Alcazar17, Prieto17, Trebitsch18}.

\subsubsection{Impact of SN Feedback Strength on BH Growth} 
\label{sec-res-Vary-SN-strength-BH-growth} 

\begin{figure*} 
\centering 
\includegraphics[width = 1.1 \linewidth]{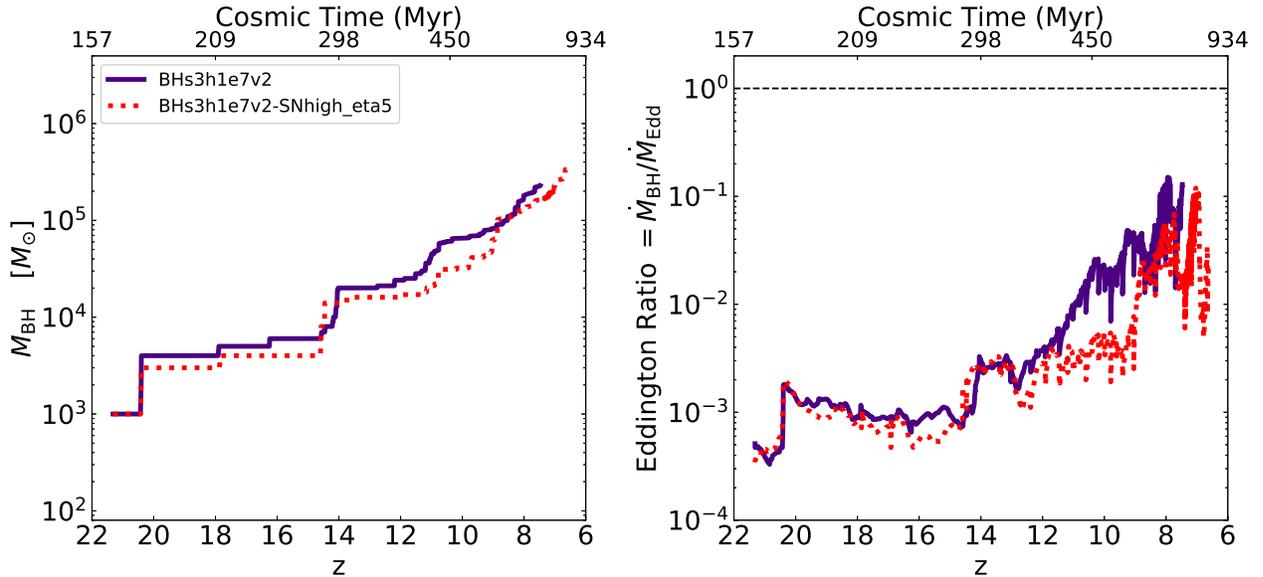} 
\caption{ 
BH mass (left panel) and BH accretion rate Eddington ratio (right panel) growth 
with redshift in two runs with different strengths ($\eta$) of SN feedback. 
The colours discriminate the runs as labelled, which are: 
{\it BHs3h1e7v2}      - indigo solid, and 
{\it BHs3h1e7v2-SNhi} - red dashed. 
} 
\label{fig-BH-Mass-and-AccRate-vs-Time} 
\end{figure*} 

The growth of the most-massive BH in runs {\it BHs3h1e7v2-SNhi} and {\it BHs3h1e7v2} 
are plotted in Fig.~\ref{fig-BH-Mass-and-AccRate-vs-Time}: 
redshift evolution of BH mass in the left panel, 
and accretion rate Eddington ratio at the right. 
The BH properties remain more-or-less the same in both simulations. 
After being seeded at $z \sim 21$ as $10^{3} M_{\odot}$ BHs, they grow in mass 
by accreting dense gas from their surroundings and by merger with other BHs, 
to $M_{\rm BH} = 3 \times 10^{5} M_{\odot}$ by $z \sim 7$. 
The accretion rate grows as well similarly from Eddington ratio $< 0.0001$ at seeding, 
to $\dot{M}_{\rm BH} \sim 0.2 \dot{M}_{\rm Edd}$ by $z \sim 7$, 
with substantial variability of the $\dot{M}_{\rm BH}$ at later epochs. 

Thus, we find that in our simulations varying SN feedback 
has a little impact on BH growth. Our BHs continue to grow up to $z \sim 6$ 
even in the presence of stronger SN-driven mass ejection. 
We note that this result is contrary to some recent zoomed-in cosmological 
hydrodynamical simulation studies 
\citep[e.g.,][]{Dubois15, Angles-Alcazar17, Prieto17, Trebitsch18} which find 
that strong SN feedback hampers BH growth, by removing gas from around the BH. 
Simulating a $10^{12} M_{\odot}$ halo, \citet{Dubois15} found that at $z > 3.5$ 
efficient feedback from SN destroys dense cold gas clumps at the galaxy core, 
preventing early BH growth. 
However later on, BH growth occurs from gas inflows induced by major mergers of galaxies. 
Simulating $10^{12.5} M_{\odot}$ haloes down to $z = 1$, 
\citet{Angles-Alcazar17} showed that early BH growth is limited by 
bursty stellar feedback evacuating gas from galactic nuclei. 
Simulating a $3 \times 10^{10} M_{\odot}$ halo down to $z = 6$, 
\citet{Prieto17} showed that, in the absence of SN feedback, 
AGN feedback alone does not affect significantly either SF or BH growth. 
Performing zoom-in radiation hydrodynamics simulations of a 
$5 \times 10^{9} M_{\odot}$ halo down to $z = 5.7$, \citet{Trebitsch18} found that 
SN feedback prevents the growth of the BH, thus quenching its associated feedback. 

In another study, \citet{Habouzit17} performed periodic $(10$ Mpc$)^3$ 
simulations up to $z = 2$, using a new implementation of BH seed 
formation in low-metallicity environments, and found that all high-mass galaxies 
tend to host a BH, while low-mass galaxies have a lower probability of hosting a BH. 
In \citet{Dubois15} and \citet{Habouzit17}, the BH growth is hampered 
by strong SN feedback, only for their delayed cooling SN feedback implementation 
where gas cooling is turned off for a certain duration. 
Thus the numerical formulation of the relevant physical processes 
can contribute to obtaining different results in the simulations. 

We remind below some features in our sub-resolution models which 
lead to the SN feedback being less coupled to the galaxy ISM. 
In our simulations, newly kicked SN wind particles are decoupled from 
hydrodynamic interactions (as described in \S\ref{sec-num-cool-SF-SN}, last paragraph). 
This is done in order to create SN outflows without destroying 
dense star forming regions. It allows SN winds to travel until reaching a density 
$\rho_{\rm dec} = 0.25 \rho_{\rm SF} = 0.03$ cm$^{-3}$ (number density 
of hydrogen atoms), before being hydrodynamically coupled again. 
Although widely used, this implementation effectively 
allows SN winds to escape the galaxy, while decoupled from the ISM. 
So even though we have increased the mass loading factor of SN kinetic feedback 
in the new run {\it BHs3h1e7v2-SNhi}, it does not necessarily mean that 
the coupling between the SN feedback and the ISM is stronger, 
because of the hydrodynamic decoupling feature present. 
Also note that our SN wind particles are kicked in a direction $+$ or $-$ 
$(\vec{v}_{\rm old} \times \vec{\nabla} \phi)$ 
(\S\ref{sec-num-cool-SF-SN}, fifth paragraph), with an aim to create 
SN outflows preferentially ejected perpendicular to the galaxy disk. 
Such a prescription reduces even more the coupling between 
SN feedback and the ISM of the galaxy. 

Given the above aspects of our SN outflow sub-resolution scheme, 
it can be unexpected to find that our star formation is strongly quenched by 
SN feedback (\S\ref{sec-res-Vary-SN-strength-SFR}), 
higher than the quenching by BH feedback. 
This can probably be explained by the low value of the star-formation 
threshold density used (\S\ref{sec-num-cool-SF-SN}, second paragraph): 
$\rho_{\rm SF} = 0.13$ cm$^{-3}$. 
As a result, even the diffuse gas in the galaxy is considered as star forming, 
and will be the most affected by SN feedback. 

Hence to summarize, the apparent discrepancy of our result (continued BH growth) 
with the other studies in the literature is because of several conditions: 
environmental effects and our cosmological setup, 
combined with the different numerical implementation 
of the sub-resolution physical processes. 
We have smaller BHs in smaller halos, but in a full cosmological environment; 
therefore the effects of galaxy mergers and hierarchical structure formation 
should be fully taken into account. 
In our simulations, the supply of gas is maintained for a longer time, 
so that BHs can grow even in the presence of stronger SN feedback. 
As galaxies assemble in a cosmological environment, there are gas inflows to 
centers driven by major/minor galaxy mergers, tidal encounters, galaxy harassment etc. 
This sustained accumulation of gas flows over longer time periods 
enable the central BHs to grow. 
Such a conclusion of late BH growth was also implied by \citet{Dubois15}. 

Earlier we discussed our finding that the accretion rate $\dot{M}_{\rm BH}$ 
of the growing BHs is highly variable 
(Fig.~\ref{fig-BH-AccRate-vs-Time} in \S\ref{sec-res-BH-Growth}, and 
Fig.~\ref{fig-BH-Mass-and-AccRate-vs-Time} - right panel in 
\S\ref{sec-res-Vary-SN-strength-BH-growth}). 
A possible reason for this phenomenon is self-regulation by the feedback 
from BHs and/or supernova. 
A cycle of accretion-induced inflow is followed by BH- and/or SN-driven 
feedback-induced outflow, which makes the gas supply intermittent. 
At the same time, gas supply is not depleted in the long term. 
The inflow-outflow cycles continue up to $z \sim 4$, and the BHs grow during inflows. 

Note that \citet{Habouzit17} also performed fully cosmological simulations, 
however found that SN feedback prevents BH growth in low-mass galaxies, 
but only when using a more efficient SN feedback sub-resolution recipe. 

In a recent study, \citet{Koudmani18} performed high-resolution isolated DG 
simulations, using a constant BH accretion rate equal to a fixed fraction 
of the Eddington rate based on the initial BH mass. 
Thus mimicking a maximal feedback effect, they found that AGN outflows 
have a small but systematic effect on the galaxy central SFRs. 
While significant effects on the global SFR are only obtained 
with strong SN feedback and sustained high-luminosity isotropic AGN winds. 
The outflows reach much higher velocities and much-higher temperatures 
with AGN feedback included. 




\section{Summary and Conclusions} 
\label{sec-conclusion} 

Intermediate-Mass Black Holes (with masses between $100 - 10^{6} M_{\odot}$) 
have historically been an elusive population of BHs, compared to the stellar-mass 
and supermassive (widely observed at the centers of AGN) BH counterparts. 
Recently these IMBHs have started to be observed in low-mass galaxies. 
Our work focuses on the case that IMBHs are formed at the centers of dwarf galaxies. 
Early feedback from such IMBHs is expected to release energy and affect the host 
gas-rich dwarf galaxies at $z = 5 - 8$, quenching star-formation, 
reducing the number of DGs, and impacting the density profile at DG centers. 
This can possibly solve several anomalies in the dwarf galaxy mass range within 
the concordance $\Lambda$CDM cosmological scenario of galaxy formation \citep{Silk17}. 

We have investigated the growth and feedback of IMBHs at DG centers, 
by performing cosmological hydrodynamical simulations. 
We have employed a modified version of the SPH code GADGET-3. 
It includes the following sub-resolution physics: 
radiative cooling and heating from a photoionizing background, 
star-formation, stellar evolution, chemical enrichment for $11$ elements, 
supernova feedback, AGN accretion and feedback. 
We simulated $(2 h^{-1} ~ {\rm Mpc})^3$ comoving volumes 
to probe dwarf galaxies at high redshifts. 
The mass resolutions are $3.44 \times 10^{4} h^{-1} M_{\odot}$ for dark matter 
particles, and $6.43 \times 10^{3} h^{-1} M_{\odot}$ for gas particles. 
The length resolution is $0.1 h^{-1}$ kpc comoving 
which is the gravitational softening length. 
The cosmological boxes are evolved with periodic boundary conditions, 
from $z = 100$ up to $z = 4$. 

We executed a series of simulations: one of them is a control case 
with SF-only and no BH; the other runs include BHs 
and explore different parameter variations of the BH sub-resolution models. 
In particular, we seed BHs of mass: 
$M_{\rm BHseed} = (10^{2}, 10^{3}, 10^{4}) M_{\odot}$, 
at the centers of massive halos with 
$M_{\rm halo} > M_{\rm HaloMin} = (10^{6} - 5 \times 10^{7}) M_{\odot}$. 
In addition to $M_{\rm HaloMin}$ and $M_{\rm BHseed}$, we also vary the 
outflow velocity for BH kinetic feedback: $v_w = (1000 - 10000)$ km/s. 

The earliest BHs appear at $z \sim 18 - 25$ in our simulations, 
when the first halos reach the corresponding lower limit $M_{\rm HaloMin}$. 
The BHs are allowed to grow by accreting surrounding gas and by merger with other BHs. 
As they accrete and grow, the BHs eject out feedback energy, and impact their host DGs. 
We find the following results for the growth and feedback of IMBHs in our simulations: 

\begin{itemize} 

\item The BHs grow to intermediate masses 
$M_{\rm BH} =$ a few$\times 10^{5} - 10^{6} M_{\odot}$ by $z = 5$, when the BHs 
are seeded in halos with $M_{\rm HaloMin} \leq 4 \times 10^{7} M_{\odot}$. 
The most-massive BH at $z = 4$ has: 
$M_{\rm BH} = 2 \times 10^6 M_{\odot}$ and $\dot{M}_{\rm BH} = 0.02 M_{\odot}$/yr. 

\item BHs seeded in smaller halos grow faster (considering the same redshift) 
than those seeded in larger halos. 
E.g. at $z = 7$, the most-massive BH in the simulation has grown to 
$M_{\rm BH} = 10^5 M_{\odot}$ when seeded in 
$M_{\rm HaloMin} = 2 \times 10^{7} M_{\odot}$ halos, 
while it grows only to $M_{\rm BH} = 4 \times 10^4 M_{\odot}$ 
when seeded in $M_{\rm HaloMin} = 3 \times 10^{7} M_{\odot}$ halos. 

\item The effect of increasing or decreasing the parameters $M_{\rm HaloMin}$ 
and $M_{\rm BHseed}$ together on the BH mass growth can be the same. 
E.g., a BH can grow similarly when seeded in smaller halos 
$(M_{\rm HaloMin} = 10^{6} M_{\odot})$ 
with a smaller seed mass $(M_{\rm BHseed} = 10^{2} M_{\odot})$, 
as in larger halos $(M_{\rm HaloMin} = 10^{7} M_{\odot})$ 
with a higher seed mass $(M_{\rm BHseed} = 10^{3} M_{\odot})$. 

The variation in the BH outflow velocity has little effect upon the BH growth, 
although the increase of the outflow velocity delays the suppression of SF. 

\item Starting from highly sub-Eddington rates (Eddington ratio $< 0.001$), 
the accretion rate of the BHs increases with time, and reaches 
$\dot{M}_{\rm BH} = (0.2 - 0.8) \dot{M}_{\rm Edd}$ for the massive IMBHs by $z = 4$. 

\item Our simulations probe dwarf galaxies with a 
stellar mass between $M_{\star} = (10^{5} - 10^{8}) M_{\odot}$. 
The star formation rate density evolution of these DGs has a wide maximum 
in the form of a plateau between $z = 4 - 6$, 
and the SFRD decreases at $z > 6$ and at $z < 4$. 

\item Star-formation is quenched between $z = 9 - 4$ by BH accretion and feedback. 
The SFRD is reduced by factors $1.1 - 3$, 
when the BHs have grown to a mass $M_{\rm BH} \sim$ a few times $10^5 M_{\odot}$. 

\item There is a huge scatter in the BH - galaxy $[M_{\rm BH} - M_{\star}]$ correlation 
of the simulated galaxies, especially at low-$M_{\star}$. 

In our runs where SF quenching happens the earliest, 
the IMBHs $(M_{\rm BH} = 10^{3} - 10^{5} M_{\odot})$ 
in DGs $(M_{\star} = 10^{4} - 10^{7} M_{\odot})$ are already more massive at 
$z = 7.9$, as compared to the local $[M_{\rm BH} - M_{\star}]$ correlation 
and that of high-$z$ quasars. 
Central BHs in some runs ({\it BHs3h3e7}, {\it BHs3h4e7}, {\it BHs3h5e7}) 
more or less follow the local $z = 0$ correlation at $z \geq 5.5$. 

\item Varying the SN feedback strength (by a factor 2.5) has a little impact 
on the growth of the BHs. 
Our simulated BHs continue to grow up to $z \sim 6$, 
even in the presence of stronger SN-driven mass ejection, sustained by gas 
inflows driven by galaxy mergers and interactions in a cosmological environment. 

\item The IMBHs do not always show indications of BH feedback driven gas outflows. 
There are only signatures of outflows around the most-massive IMBHs 
with $M_{\rm BH} > 10^6 M_{\odot}$, as shock-heated low-density gas. 

\item Our result of rapid BH growth at the centers of DGs suggest that 
these BHs grow faster than their host galaxies in the early Universe. 
The resulting early BH feedback quenches star formation, 
and possibly turns the dwarf galaxies (as well as their central BHs) dormant. 

\end{itemize} 

{\it We deduce that intermediate-mass black holes at the centers 
of dwarf galaxies can be a strong source of feedback. 
Our cosmological hydrodynamical simulations show that when the IMBHs have grown to 
$10^5 M_{\odot}$, they quench star formation in their host galaxies. 
At the same time, these IMBHs form the missing link between stellar-mass 
and supermassive BHs.}

\section*{Acknowledgments} 


We are most grateful to Volker Springel for allowing us to use the GADGET-3 code. 
This work is supported by the Brazilian Agencies 
FAPESP (grants 2016/01355-5, 2016/22183-8, and 2013/10559-5); 
and CNPq (grant 308643/2017-8).

%



\begin{thebibliography}{99} 

\bibitem[\protect\citeauthoryear{Afanasiev et al.}{2018}]{Afanasiev18} 
Afanasiev, A. V. et al. 2018, accepted in MNRAS, eprint arXiv:1804.02938 

\bibitem[\protect\citeauthoryear{Ahn et al.}{2018}]{Ahn18} 
Ahn, C. P. et al. 2018, eprint arXiv:1804.02399 

\bibitem[\protect\citeauthoryear{Angles-Alcazar et al.}{2017}]{Angles-Alcazar17} 
Angles-Alcazar, D., Faucher-Giguere, C.-A., Quataert, E., Hopkins, P. F., 
Feldmann, R., Torrey, P., Wetzel, A. \& Keres, D. 2017, MNRAS, 472, L109 

\bibitem[\protect\citeauthoryear{Baldassare et al.}{2015}]{Baldassare15} 
Baldassare, V. F., Reines, A. E., Gallo, E. \& Greene, J. E. 2015, ApJ, 809, L14 

\bibitem[\protect\citeauthoryear{Barai}{2008}]{Barai08} 
Barai, P. 2008, ApJ, 682, L17 

\bibitem[\protect\citeauthoryear{Barai et al.}{2013}]{Barai13} 
Barai, P. et al. 2013, MNRAS, 430, 3213 

\bibitem[\protect\citeauthoryear{Barai et al.}{2014}]{Barai14} 
Barai, P., Viel, M., Murante, G., Gaspari, M. \& Borgani, S. 2014, MNRAS, 437, 1456 

\bibitem[\protect\citeauthoryear{Barai et al.}{2015}]{Barai15} 
Barai, P., Monaco, P., Murante, G., Ragagnin, A. \& Viel, M. 2015, MNRAS, 447, 266 

\bibitem[\protect\citeauthoryear{Barai et al.}{2016}]{Barai16} 
Barai, P., Murante, G., Borgani, S., Gaspari, M., Granato, G. L., Monaco, P. \& Ragone-Figueroa, C. 2016, MNRAS, 461, 1548 

\bibitem[\protect\citeauthoryear{Barai et al.}{2018}]{Barai18} 
Barai, P., Gallerani, S., Pallottini, A., Ferrara, A., Marconi, A., Cicone, C., 
Maiolino, R. \& Carniani, S. 2018, MNRAS, 473, 4003 

\bibitem[\protect\citeauthoryear{Bieri et al.}{2015}]{Bieri15} 
Bieri, R., Dubois, Y., Silk, J. \& Mamon, G. A. 2015, ApJ, 812, L36 

\bibitem[\protect\citeauthoryear{Biffi et al.}{2016}]{Biffi16} 
Biffi, V., Borgani, S., Murante, G., Rasia, E., Planelles, S., Granato, G. L., 
Ragone-Figueroa, C., Beck, A. M., Gaspari, M. \& Dolag, K. 2016, ApJ, 827, 112 

\bibitem[\protect\citeauthoryear{Bolton}{1972}]{Bolton72} 
Bolton, C. T. 1972, Nature Physical Science, 240, 124 

\bibitem[\protect\citeauthoryear{Bondi}{1952}]{Bondi52} 
Bondi, H. 1952, MNRAS, 112, 195 

\bibitem[\protect\citeauthoryear{Bouche et al.}{2012}]{Bouche12} 
Bouche, N., Hohensee, W., Vargas, R., Kacprzak, G. G., Martin, C. L., Cooke, J. 
\& Churchill, C. W. 2012, MNRAS, 426, 801 

\bibitem[\protect\citeauthoryear{Bouwens et al.}{2009}]{Bouwens09} 
Bouwens, R. J. et al. 2009, ApJ, 705, 936 

\bibitem[\protect\citeauthoryear{Bouwens et al.}{2012}]{Bouwens12} 
Bouwens, R. J. et al. 2012, ApJ, 754, 83 

\bibitem[\protect\citeauthoryear{Caballero-Garcia et al.}{2018}]{Caballero-Garcia18} 
Caballero-Garcia, M. D., Fabrika, S., Castro-Tirado, A. J., Bursa, M., Dovciak, M., 
Castellon, A. \& Karas, V. 2018, submitted to RMxAC, eprint arXiv:1802.07149 

\bibitem[\protect\citeauthoryear{Caproni et al.}{2017}]{Caproni17} 
Caproni, A., Amaral Lanfranchi, G., Campos Baio, G. H., Kowal, G., 
\& Falceta-Goncalves, D. 2017, ApJ, 838, 99 

\bibitem[\protect\citeauthoryear{Chabrier}{2003}]{Chabrier03} 
Chabrier, G. 2003, PASP, 115, 763 

\bibitem[\protect\citeauthoryear{Chambers, Miley \& van Breugel}{1987}]{Chambers87} 
Chambers, K. C., Miley, G. K. \& van Breugel, W. 1987, Nature, 329, 604 

\bibitem[\protect\citeauthoryear{Chilingarian et al.}{2018}]{Chilingarian18} 
Chilingarian, I. V., Katkov, I. Y., Zolotukhin, I. Y., Grishin, K. A., Beletsky, Y., 
Boutsia, K. \& Osip, D. J. 2018, submitted to ApJ, eprint arXiv:1805.01467 

\bibitem[\protect\citeauthoryear{Cicone et al.}{2014}]{Cicone14} 
Cicone, C. et al. 2014, A\&A, 562, A21 

\bibitem[\protect\citeauthoryear{Corral-Santana et al.}{2016}]{Corral16} 
Corral-Santana, J. M., Casares, J., Munoz-Darias, T., Bauer, F. E., 
Martinez-Pais, I. G. \& Russell, D. M. 2016, A\&A, 587, A61 

\bibitem[\protect\citeauthoryear{Cowley et al.}{1983}]{Cowley83} 
Cowley, A. P., Crampton, D., Hutchings, J. B., Remillard, R. 
\& Penfold, J. E. 1983, ApJ, 272, 118 

\bibitem[\protect\citeauthoryear{Crenshaw, Kraemer \& George}{2003}]{Crenshaw03} 
Crenshaw, D. M., Kraemer, S. B., \& George, I. M. 2003, ARA\&A, 41, 117 

\bibitem[\protect\citeauthoryear{Cucciati et al.}{2012}]{Cucciati12} 
Cucciati, O. et al. 2012, A\&A, 539, A31 

\bibitem[\protect\citeauthoryear{Dashyan et al.}{2018}]{Dashyan17} 
Dashyan, G., Silk, J., Mamon, G. A., Dubois, Y. \& Hartwig, T. 2018, MNRAS, 473, 5698 

\bibitem[\protect\citeauthoryear{Demers et al.}{1995}]{Demers95} 
Demers, S., Battinelli, P., Irwin, M. J. \& Kunkel, W. E. 1995, MNRAS, 274, 491 

\bibitem[\protect\citeauthoryear{De Young}{1989}]{DeYoung89} 
De Young, D. S. 1989, ApJ, 342, L59 

\bibitem[\protect\citeauthoryear{Di Matteo et al.}{2012}]{DiMatteo12} 
Di Matteo, T., Khandai, N., DeGraf, C., Feng, Y., Croft, R. A. C., 
Lopez, J. \& Springel, V. 2012, ApJ, 745, L29 

\bibitem[\protect\citeauthoryear{Dubois et al.}{2013}]{Dubois13} 
Dubois, Y., Pichon, C., Devriendt, J., Silk, J., Haehnelt, M., Kimm, T. 
\& Slyz, A. 2013, MNRAS, 428, 2885 

\bibitem[\protect\citeauthoryear{Dubois et al.}{2015}]{Dubois15} 
Dubois, Y., Volonteri, M., Silk, J., Devriendt, J., Slyz, A. \& Teyssier, R. 
2015, MNRAS, 452, 1502 

\bibitem[\protect\citeauthoryear{Fabian}{2012}]{Fabian12} 
Fabian, A. C. 2012, ARA\&A, 50, 455 

\bibitem[\protect\citeauthoryear{Ferland et al.}{1998}]{Ferland98} 
Ferland, G. J., Korista, K. T., Verner, D. A., Ferguson, J. W., 
Kingdon, J. B. \& Verner, E. M. 1998, PASP, 110, 761 

\bibitem[\protect\citeauthoryear{Ferrarese \& Ford}{2005}]{Ferrarese05} 
Ferrarese, L., \& Ford, H. 2005, SSRv, 116, 523 

\bibitem[\protect\citeauthoryear{Filippenko \& Sargent}{1989}]{Filippenko89} 
Filippenko, A. V. \& Sargent, W. L. W. 1989, ApJ, 342, L11 

\bibitem[\protect\citeauthoryear{Gebhardt et al.}{2000}]{Gebhardt00} 
Gebhardt, K. et al. 2000, ApJ, 539, L13 

\bibitem[\protect\citeauthoryear{Giesers et al.}{2018}]{Giesers18} 
Giesers, B. et al. 2018, MNRAS, 475, L15 

\bibitem[\protect\citeauthoryear{Goulding \& Alexander}{2009}]{Goulding09} 
Goulding, A. D. \& Alexander, D. M. 2009, MNRAS, 398, 1165 

\bibitem[\protect\citeauthoryear{Graham et al.}{2018}]{Graham18} 
Graham, A. W., Soria, R. \& Davis, B. L. 2018, MNRAS accepted, 
eprint arXiv:1811.03232 

\bibitem[\protect\citeauthoryear{Haardt \& Madau}{2001}]{Haardt01} 
Haardt, F. \& Madau, P. 2001, XXXVIth Rencontres de Moriond, 
XXIst Moriond Astrophysics Meeting, Editors D.M.Neumann \& J.T.T.Van, 64 

\bibitem[\protect\citeauthoryear{Habouzit et al.}{2017}]{Habouzit17} 
Habouzit, M., Volonteri, M. \& Dubois, Y. 2017, MNRAS, 468, 3935 

\bibitem[\protect\citeauthoryear{Hahn \& Abel}{2011}]{Hahn11} 
Hahn, O. \& Abel, T. 2011, MNRAS, 415, 2101 

\bibitem[\protect\citeauthoryear{Hoyle \& Lyttleton}{1939}]{Hoyle39} 
Hoyle, F. \& Lyttleton, R. A. 1939, Proc. Cam. Phil. Soc., 35, 405 

\bibitem[\protect\citeauthoryear{Izotov \& Thuan}{2008}]{Izotov08} 
Izotov, Y. I. \& Thuan, T. X. 2008, ApJ, 687, 133 

\bibitem[\protect\citeauthoryear{Johansson, Naab \& Burkert}{2009}]{Johansson09a} 
Johansson, P. H., Naab, T. \& Burkert, A. 2009, ApJ, 690, 802 

\bibitem[\protect\citeauthoryear{Katz, Weinberg \& Hernquist}{1996}]{Katz96} 
Katz, N., Weinberg, D. H. \& Hernquist, L. 1996, ApJS, 105, 19 

\bibitem[\protect\citeauthoryear{Kennicutt}{1998}]{Kennicutt98} 
Kennicutt, R. C. 1998, ApJ, 498, 541 

\bibitem[\protect\citeauthoryear{Koudmani et al.}{2018}]{Koudmani18} 
Koudmani, S., Sijacki, D., Bourne, M. A. \& Smith, M. C. 2019, MNRAS, 484, 2047 

\bibitem[\protect\citeauthoryear{Kovetz et al.}{2018}]{Kovetz18} 
Kovetz, E. D., Cholis, I., Kamionkowski, M. \& Silk, J. 2018, eprint arXiv:1803.00568 

\bibitem[\protect\citeauthoryear{Kraemer, Tombesi \& Bottorff}{2018}]{Kraemer18} 
Kraemer, S. B., Tombesi, F. \& Bottorff, M. C. 2018, ApJ, 852, 35 

\bibitem[\protect\citeauthoryear{Kunth, Sargent \& Bothun}{1987}]{Kunth87} 
Kunth, D., Sargent, W. L. W. \& Bothun, G. D. 1987, AJ, 93, 29 

\bibitem[\protect\citeauthoryear{Lanz et al.}{2016}]{Lanz16} 
Lanz, L., Ogle, P. M., Alatalo, K. \& Appleton, P. N. 2016, ApJ, 826, 29 

\bibitem[\protect\citeauthoryear{Lemons et al.}{2015}]{Lemons15} 
Lemons, S. M., Reines, A. E., Plotkin, R. M., Gallo, E. 
\& Greene, J. E. 2015, ApJ, 805, 12 

\bibitem[\protect\citeauthoryear{Maccarone et al.}{2007}]{Maccarone07} 
Maccarone, T. J., Kundu, A., Zepf, S. E. \& Rhode, K. L. 2007, Natur, 445, 183 

\bibitem[\protect\citeauthoryear{Magorrian et al.}{1998}]{Magorrian98} 
Magorrian, J. et al. 1998, AJ, 115, 2285 

\bibitem[\protect\citeauthoryear{Marconi \& Hunt}{2003}]{Marconi03} 
Marconi, A. \& Hunt, L. K. 2003, ApJ, 589, L21 

\bibitem[\protect\citeauthoryear{Marleau et al.}{2017}]{Marleau17} 
Marleau, F. R., Clancy, D., Habas, R. \& Bianconi, M. 2017, A\&A, 602, A28 

\bibitem[\protect\citeauthoryear{Martin}{1999}]{Martin99} 
Martin, C. L. 1999, ApJ, 513, 156 

\bibitem[\protect\citeauthoryear{Martin-Navarro \& Mezcua}{2018}]{Martin-Navarro18} 
Martin-Navarro, I. \& Mezcua, M. 2018, ApJ, 855, L20 

\bibitem[\protect\citeauthoryear{McConnell \& Ma}{2013}]{McConnell13} 
McConnell, N. J. \& Ma, C.-P. 2013, ApJ, 764, 184 

\bibitem[\protect\citeauthoryear{Melioli, de Gouveia Dal Pino \& Geraissate}{2013}]{Melioli13} 
Melioli, C., de Gouveia Dal Pino, E. M. \& Geraissate, F. G. 2013, MNRAS, 430, 3235 

\bibitem[\protect\citeauthoryear{Melioli \& de Gouveia Dal Pino}{2015}]{Melioli15} 
Melioli, C. \& de Gouveia Dal Pino, E. M. 2015, ApJ, 812, 90 

\bibitem[\protect\citeauthoryear{Mezcua et al.}{2018a}]{Mezcua18a} 
Mezcua, M., Civano, F., Marchesi, S., Suh, H., Fabbiano, G. \& Volonteri, M. 
2018a, MNRAS, 478, 2576 

\bibitem[\protect\citeauthoryear{Mezcua et al.}{2018b}]{Mezcua18b} 
Mezcua, M., Kim, M., Ho, L. C. \& Lonsdale, C. J. 2018b, MNRAS, 480, L74 

\bibitem[\protect\citeauthoryear{Miller et al.}{2003}]{Miller03} 
Miller, J. M., Fabbiano, G., Miller, M. C. \& Fabian, A. C. 2003, ApJ, 585, L37 

\bibitem[\protect\citeauthoryear{Moran et al.}{2014}]{Moran14} 
Moran, E. C., Shahinyan, K., Sugarman, H. R., Velez, D. O. 
\& Eracleous, M. 2014, AJ, 148, 136 

\bibitem[\protect\citeauthoryear{Mortlock et al.}{2011}]{Mortlock11} 
Mortlock, D. J. et al. 2011, Nature, 474, 616 

\bibitem[\protect\citeauthoryear{Mukherjee et al.}{2018}]{Mukherjee18} 
Mukherjee, D., Bicknell, G. V., Wagner, A. Y., Sutherland, R. S. \& 
Silk, J. 2018, MNRAS, 479, 5544 

\bibitem[\protect\citeauthoryear{Orosz et al.}{2007}]{Orosz07} 
Orosz, J. A. et al. 2007, Nature, 449, 872 

\bibitem[\protect\citeauthoryear{Penny et al.}{2017}]{Penny17} 
Penny, S. J. et al. 2017, submitted to MNRAS, eprint arXiv:1710.07568 

\bibitem[\protect\citeauthoryear{Perez-Gonzalez et al.}{2005}]{PerezGonzalez05} 
Perez-Gonzalez, P. G. et al. 2005, ApJ, 630, 82 

\bibitem[\protect\citeauthoryear{Peterson et al.}{2005}]{Peterson05} 
Peterson, B. M. et al. 2005, ApJ, 632, 799 

\bibitem[\protect\citeauthoryear{Planck Collaboration}{2015}]{Planck15} 
Planck Collaboration; Ade, P. A. R. et al. 2016, A\&A, 594, A13 

\bibitem[\protect\citeauthoryear{Prieto et al.}{2017}]{Prieto17} 
Prieto, J., Escala, A., Volonteri, M. \& Dubois, Y. 2017, ApJ, 836, 216 

\bibitem[\protect\citeauthoryear{Reddy \& Steidel}{2009}]{Reddy09} 
Reddy, N. A. \& Steidel, C. C. 2009, ApJ, 692, 778 

\bibitem[\protect\citeauthoryear{Rees}{1984}]{Rees84} 
Rees, M. J. 1984, ARA\&A, 22, 471 

\bibitem[\protect\citeauthoryear{Reines et al.}{2011}]{Reines11} 
Reines, A. E., Sivakoff, G. R., Johnson, K. E. \& Brogan, C. L. 2011, Nature, 470, 66 

\bibitem[\protect\citeauthoryear{Reines, Greene \& Geha}{2013}]{Reines13} 
Reines, A. E., Greene, J, E. \& Geha, M. 2013, ApJ, 775, 116 

\bibitem[\protect\citeauthoryear{Reines \& Volonteri}{2015}]{Reines15} 
Reines, A. E. \& Volonteri, M. 2015, ApJ, 813, 82 

\bibitem[\protect\citeauthoryear{Rodighiero et al.}{2010}]{Rodighiero10} 
Rodighiero, G. et al. 2010, A\&A, 515, A8 

\bibitem[\protect\citeauthoryear{Salmon et al.}{2015}]{Salmon15} 
Salmon, B. et al. 2015, ApJ, 799, 183 

\bibitem[\protect\citeauthoryear{Salome et al.}{2017}]{Salome17} 
Salome, Q., Salome, P., Miville-Deschenes, M.-A., Combes, F. \& 
Hamer, S. 2017, A\&A, 608, A98 

\bibitem[\protect\citeauthoryear{Santini et al.}{2017}]{Santini17} 
Santini, P. et al. 2017, ApJ, 847, 76 

\bibitem[\protect\citeauthoryear{Scannapieco, Silk \& Bouwens}{2005}]{Scannapieco05} 
Scannapieco, E., Silk, J. \& Bouwens, R. 2005, ApJ, 635, L13 

\bibitem[\protect\citeauthoryear{Schawinski et al.}{2006}]{Schawinski06} 
Schawinski, K. et al. 2006, Nature, 442, 888 

\bibitem[\protect\citeauthoryear{Schaye et al.}{2015}]{Schaye15} 
Schaye, J. et al. 2015, MNRAS, 446, 521 

\bibitem[\protect\citeauthoryear{Schiminovich et al.}{2005}]{Schiminovich05} 
Schiminovich, D. et al. 2005, ApJ, 619, L47 

\bibitem[\protect\citeauthoryear{Schmidt}{1959}]{Schmidt59} 
Schmidt, M. 1959, ApJ, 129, 243 

\bibitem[\protect\citeauthoryear{Schramm et al.}{2013}]{Schramm13} 
Schramm, M. et al. 2013, ApJ, 773, 150 

\bibitem[\protect\citeauthoryear{Secrest et al.}{2015}]{Secrest15} 
Secrest, N. J. et al. 2015, ApJ, 798, 38 

\bibitem[\protect\citeauthoryear{Seth et al.}{2010}]{Seth10} 
Seth, A. C. et al. 2010, ApJ, 714, 713 

\bibitem[\protect\citeauthoryear{Shakura \& Sunyaev}{1973}]{Shakura73} 
Shakura, N. I. \& Sunyaev, R. A. 1973, A\&A, 24, 337 

\bibitem[\protect\citeauthoryear{Sijacki et al.}{2007}]{Sijacki07} 
Sijacki, D., Springel, V., Di Matteo, T. \& Hernquist, L. 2007, MNRAS, 380, 877 

\bibitem[\protect\citeauthoryear{Silk \& Rees}{1998}]{Silk98} 
Silk, J., \& Rees, M. J. 1998, A\&A, 331, L1 

\bibitem[\protect\citeauthoryear{Silk}{2017}]{Silk17} 
Silk, J. 2017, ApJ, 839, L13 

\bibitem[\protect\citeauthoryear{Springel \& Hernquist}{2003}]{SH03} 
Springel, V. \& Hernquist, L. 2003, MNRAS, 339, 289 

\bibitem[\protect\citeauthoryear{Springel}{2005}]{Springel05} 
Springel, V. 2005, MNRAS, 364, 1105 

\bibitem[\protect\citeauthoryear{Springel, Di Matteo \& Hernquist}{2005}]{SDH05} 
Springel, V., Di Matteo, T. \& Hernquist, L. 2005, MNRAS, 361, 776 

\bibitem[\protect\citeauthoryear{Sutton et al.}{2012}]{Sutton12} 
Sutton, A. D., Roberts, T. P., Walton, D. J., Gladstone, J. C. 
\& Scott, A. E. 2012, MNRAS, 423, 1154 

\bibitem[\protect\citeauthoryear{Tescari et al.}{2011}]{Tescari11} 
Tescari, E., Viel, M., D'Odorico, V., Cristiani, S., Calura, F., 
Borgani, S. \& Tornatore, L. 2011, MNRAS, 411, 826 

\bibitem[\protect\citeauthoryear{Thornton et al.}{2008}]{Thornton08} 
Thornton, C. E., Barth, A. J., Ho, L. C., Rutledge, R. E. 
\& Greene, J. E. 2008, ApJ, 686, 892 

\bibitem[\protect\citeauthoryear{Tombesi et al.}{2015}]{Tombesi15} 
Tombesi, F., Melendez, M., Veilleux, S., Reeves, J. N., Gonzalez-Alfonso, E. 
\& Reynolds, C. S. 2015, Nature, 519, 436 

\bibitem[\protect\citeauthoryear{Tornatore et al.}{2007}]{Tornatore07} 
Tornatore, L., Borgani, S., Dolag, K. \& Matteucci, F. 2007, MNRAS, 382, 1050 

\bibitem[\protect\citeauthoryear{Trebitsch et al.}{2018}]{Trebitsch18} 
Trebitsch, M., Volonteri, M., Dubois, Y. \& Madau, P. 2018, MNRAS, 478, 5607 

\bibitem[\protect\citeauthoryear{Tremmel et al.}{2015}]{Tremmel15} 
Tremmel, M., Governato, F., Volonteri, M. \& Quinn, T. R. 2015, MNRAS, 451, 1868 

\bibitem[\protect\citeauthoryear{Urry \& Padovani}{1995}]{Urry95} 
Urry, C. M. \& Padovani, P. 1995, PASP, 107, 803 

\bibitem[\protect\citeauthoryear{Valluri et al.}{2005}]{Valluri05} 
Valluri, M., Ferrarese, L., Merritt, D. \& Joseph, C. L. 2005, ApJ, 628, 137 

\bibitem[\protect\citeauthoryear{van der Burg, Hildebrandt \& Erben}{2010}]{vanderBurg10} 
van der Burg, R. F. J., Hildebrandt, H. \& Erben, T. 2010, A\&A, 523, A74 

\bibitem[\protect\citeauthoryear{van der Marel et al.}{1998}]{vanderMarel98} 
van der Marel, R. P., Cretton, N., de Zeeuw, P. T. \& Rix, H.-W. 1998, ApJ, 493, 613 

\bibitem[\protect\citeauthoryear{van der Marel}{2004}]{vanderMarel04} 
van der Marel, R. P. 2004, Carnegie Observatories Astrophysics Series. 
eds Ho, L. C. CUP, 37 

\bibitem[\protect\citeauthoryear{Viegas \& de Gouveia dal Pino}{1992}]{Viegas92} 
Viegas, S. M. \& de Gouveia dal Pino, E. M. 1992, ApJ, 384, 467 

\bibitem[\protect\citeauthoryear{Volonteri \& Natarajan}{2009}]{Volonteri09} 
Volonteri, M. \& Natarajan, P. 2009, MNRAS, 400, 1911 

\bibitem[\protect\citeauthoryear{Wang et al.}{2010}]{Wang10} 
Wang, R. et al. 2010, ApJ, 714, 699 

\bibitem[\protect\citeauthoryear{Wiersma, Schaye \& Smith}{2009}]{Wiersma09a} 
Wiersma, R. P. C., Schaye, J. \& Smith, B. D. 2009, MNRAS, 393, 99 

\bibitem[\protect\citeauthoryear{Woo et al.}{2019}]{Woo19} 
Woo, J.-H., Cho, H., Gallo, E., Hodges-Kluck, E., Le, H. A., Shin, J., 
Son, D. \& Horst, J. C. 2019, eprint arXiv:1905.00145, in press 

\bibitem[\protect\citeauthoryear{Wurster \& Thacker}{2013}]{Wurster13} 
Wurster, J. \& Thacker, R. J. 2013, MNRAS, 431, 2513 

\bibitem[\protect\citeauthoryear{Zinn et al.}{2013}]{Zinn13} 
Zinn, P.-C., Middelberg, E., Norris, R. P. \& Dettmar, R.-J. 2013, ApJ, 774, 66 


\end{thebibliography}
\end{document}